\documentclass[prb,aps,amssymb,twocolumn,superscriptaddress,notitlepage]{revtex4-2}
\usepackage{amsmath}
\usepackage{amssymb}
\usepackage{amsthm}
\usepackage{amsfonts}
\usepackage{listings}
\usepackage{physics}
\usepackage{tikz}
\usetikzlibrary{shapes.geometric}
\usepackage{enumerate}
\usepackage{latexsym}
\usepackage{chemfig}
\usepackage{psfrag}
\usepackage{bm}
\usepackage{array}
\usepackage{graphicx}
\usepackage{blkarray}
\usepackage{tabularx}
\usetikzlibrary{arrows.meta}
\usepackage{array}
\usepackage{color}
\usepackage{comment}
\usepackage[normalem]{ulem}
\usepackage{hyperref}
\hypersetup{
     colorlinks = true,
     linkcolor = magenta,
     citecolor = magenta
}

\def\beq{\begin{equation}}
\def\eeq{\end{equation}}
\def\bal{\begin{aligned}}
\def\eal{\end{aligned}}

\begin{document}

\title{Computing the $\mathbb{Z}_2$ Invariant in Two-Dimensional Strongly-Correlated Systems}

\author{Sounak Sinha}
\affiliation{Department of Physics, University of Illinois Urbana-Champaign, Urbana IL 61801, USA}
\affiliation{Anthony J. 
Leggett Institute for Condensed Matter Theory, University of Illinois Urbana-Champaign, Urbana IL 61801, USA}

\author{Derek Y. Pan}
\affiliation{Department of Physics, University of Illinois Urbana-Champaign, Urbana IL 61801, USA}
\affiliation{Anthony J. 
Leggett Institute for Condensed Matter Theory, University of Illinois Urbana-Champaign, Urbana IL 61801, USA}

\author{Barry Bradlyn}
\email{bbradlyn@illinois.edu}
\affiliation{Department of Physics, University of Illinois Urbana-Champaign, Urbana IL 61801, USA}
\affiliation{Anthony J. 
Leggett Institute for Condensed Matter Theory, University of Illinois Urbana-Champaign, Urbana IL 61801, USA}

\date{\today}

\begin{abstract}
We show that the two-dimensional $\mathbb{Z}_2$ invariant for time-reversal invariant insulators can be formulated in terms of the boundary-condition dependence of the ground state wavefunction for both non-interacting and strongly-correlated insulators. 
By introducing a family of quasi-single particle states associated to the many-body ground state of an insulator, we show that the $\mathbb{Z}_2$ invariant can be expressed as the integral of a certain Berry connection over half the space of boundary conditions, providing an alternative expression to the formulations that appear in [Lee et al., Phys. 
Rev. 
Lett. {\bf 100}, 186807 (2008)]. 
We show the equivalence of the different many-body formulations of the invariant, and show how they reduce to known band-theoretic results for Slater determinant ground states. 
Finally, we apply our results to analytically calculate the invariant for the Kane-Mele model with nonlocal (orbital) Hatsugai-Kohmoto (HK) interactions. 
This rigorously establishes the topological nontriviality of the Kane-Mele model with HK interactions, and represents one of the few exact calculations of the $\mathbb{Z}_2$ invariant for a strongly-interacting system.
\end{abstract}
\maketitle

\section{Introduction}

The interplay between topology and strong correlations is one of the central puzzles of modern condensed matter physics. 
On one hand, recent progress in the classification~\cite{kitaev2009periodic,chiu2016classification,kruthoff2017topological,po2017symmetrybased,bradlyn2017topological,cano2021band,cano2018building,elcoro2021magnetic,khalaf2018symmetry,po2020symmetry,hasan2010colloquium,freed2013twisted}, identification~\cite{wieder2022topological,vergniory2019complete,vergniory2022all,song2018quantitative,tang2019efficient,tang2019comprehensive}, and experimental measurement~\cite{hsieh2009observation,xu2012observation,hsu2019purely,hsu2019topology,hsieh2012topological} of topological insulators yields a relatively complete understanding of band topology and its importance to weakly correlated materials. 
Efficient techniques exist to compute topological invariants in weakly correlated systems, either directly from the ground state wave function~\cite{c.l.kaneZ_2TopologicalOrder,fuTimeReversalPolarization2006,yu2011equivalent,shiozaki2023discrete,gresch2017z2packa}, or indirectly via symmetry-based indicators~\cite{song2018quantitative,po2017symmetrybased,bradlyn2017topological,elcoro2021magnetic,khalaf2018symmetry,kruthoff2017topological}. 
At the same time, the zoo of topological phases in strongly-correlated systems is much richer, featuring phases characterized by fractionalized excitations, surface anomalies, and nontrivial long-range entanglement~\cite{wen2017colloquium,wu2012zoology,huang2017building,song2019topological,kitaev2006anyons,moore1991nonabelions,jain2007composite,thorngren2018gauging,huang2017building,barkeshli2019symmetry}. 
While effective field theory, parton-based methods, and trial wavefunction approaches offer tools for computing properties of strongly-correlated topological phases, a direct connection between topology and the microscopic physics of strongly-correlated electrons is, in general, elusive. 

With the advent of moir\'{e} materials, the connection between band topology and strongly-correlated physics has been brought to the forefront~\cite{cao2018unconventional}. 
Recent experiments in twisted multilayer graphene and twisted MoTe$_2$ have realized a variety of strongly correlated (fractional) Chern insulating and (fractional) quantum spin Hall phases~\cite{xu2023observation,lu2024fractional,cai2023signatures,park2023observation,kang2024evidence,li2021quantum,wang2020correlated}. 
Additionally, theoretical analysis suggests that the band topology (and geometry) of the non-interacting band structure in the moir\'{e} Brillouin zone plays a crucial role in stabilizing these exotic phases~\cite{ledwith2023vortexability,estienne2023ideal,andrews2024stability,roy2014band,jackson2015geometric}. 
In order to study the role of band topology in these strongly-correlated phases more directly, we need to have analytical methods for computing topological invariants for strongly-correlated systems.

For Chern numbers in (fractional) Chern insulators, the pioneering work of Niu, Thouless, and Wu (NTW) derived a formula for computing the Chern number from the ground state wave function defined with different twisted boundary conditions~\cite{niuQuantizedHallConductance1985}. 
In particular, we start by defining the ground state wave function $\ket{\psi_0(\pmb{\alpha})}$ with (magnetic) periodic boundary conditions parametrized by a vector $\pmb{\alpha}$ defined on a torus (which we call the flux Brillouin zone). 
NTW showed via the Kubo formula that the Chern number of the ground state is given by integrating the Berry curvature associated with deformations of $\pmb{\alpha}$ over the flux Brillouin zone. 
In the case of topological ground state degeneracy, one needs to carefully account for the fact that advancing $\pmb{\alpha}$ by a flux Brillouin zone lattice vector permutes the degenerate ground states; nevertheless it is possible to use the NTW formula to compute fractional Chern numbers as well.

However, for strongly-correlated quantum spin Hall systems, many-body formulations of the two-dimensional $\mathbb{Z}_2$ invariant are less well understood. 
Ref.~\cite{kaneTopologicalOrderQuantum2005} proposed a many-body extension of the Pfaffian invariant to the flux Brillouin zone, but to our knowledge no computation has been carried out using their formula. 
Along similar lines, Ref.~\cite{sheng2006quantum} introduced a many-body formulation of the spin-Chern number by using spin-resolved twisted boundary conditions, extending the NTW formula; however the spin Chern numbers are only integer-valued ground state topological invariants when a component of spin is conserved. 
Building on this, Ref.~\cite{mai2024incipient} recently computed the $\mathbb{Z}_2$ invariant of the Kane-Mele-Hubbard model with conserved $z$-component of the spin using a spin-resolved generalization of the Streda formula. 
Using effective field theory techniques, it has also been shown that the two-dimensional $\mathbb{Z}_2$ invariant is related to the partition function for a many-body system evaluated in Euclidean time on nonorientable spacetime manifolds~\cite{witten2016fermion,shiozaki2018many}. 
Similarly, Ref.~\cite{leeManyBodyGeneralizationTopological2008} introduced a many-body formulation of the two-dimensional $\mathbb{Z}_2$ invariant based on the sewing matrix for time-reversal symmetry at high-symmetry points in the flux Brillouin zone, computed in a family of excited states obtained by annihilating electrons from the ground state $\ket{\psi_0(\pmb{\alpha})}$. 
However, the relationship of this formula to the approach of Refs.~\cite{sheng2006quantum} and \cite{niuQuantizedHallConductance1985} has not been thoroughly explored.

In this work, we extend and adapt the approach of Ref.~\cite{leeManyBodyGeneralizationTopological2008} to derive several equivalent formulas for the two-dimensional $\mathbb{Z}_2$ invariant applicable to many-body systems. 
In particular, we show how the many-body $\mathbb{Z}_2$ invariant can be obtained from an integral of a particular Berry curvature and connection over half of the flux Brillouin zone, providing a many-body generalization of the alternative forms of the single-particle invariant from Ref.~\cite{fuTimeReversalPolarization2006}. 
We show that for non-interacting systems with Slater determinant ground states, our formulations of the invariant reduce to the single particle formulations of Ref.~\cite{fuTimeReversalPolarization2006}. 

Crucially, however, we go beyond previous works and apply our formulas to compute the $\mathbb{Z}_2$ invariant for the ground states of two-dimensional orbital Hatsugai-Kohmoto (HK) models~\cite{manning2023ground,jablonowski2023topological}. 
(Orbital) HK models have emerged as exactly solvable models of strongly-correlated systems with interactions diagonal in momentum space~\cite{mai20231,hatsugai1992exactly}. 
Orbital HK models captures the essential features of Mott insulators and provides exactly solvable models for topological Mott insulators with nontrivial Chern and spin Chern numbers~\cite{zhao_failure_2023,mai2024topological}. 
Furthermore, recent work has shown that the Mott transition in orbital HK models is in the same universality class as the transition in the Hubbard model, and that by considering position-space supercells, orbital HK models provide a family of exactly solvable controlled approximations to the Hubbard model~\cite{zhao2023proof,mai2024new}. 
In this work we compute the $\mathbb{Z}_2$ invariant for the Kane-Mele model in the presence of the HK interaction directly from the ground state wave function using our formulas. 
This demonstrates the utility of our approach and confirms the results of indirect calculations of topological invariants in the Kane-Mele-HK model.

To carry this out, we first in section \ref{TBC} review the formulation of the many-body NTW formula for the averaged DC conductivity. 
Next, in section \ref{many-body z2} we extend the formulation of the many-body $\mathbb{Z}_2$ invariant in Ref. \cite{leeManyBodyGeneralizationTopological2008} to a form that can be written as an obstruction to Stokes's theorem over half the flux Brillouin zone. 
Finally, in section \ref{HK}, we apply the NTW formula and the formula for the many-body $\mathbb{Z}_2$ invariant to HK models and show that there is a topological phase transitions as a function of the interaction strength $U$. 
We conclude in Sec.~\ref{sec:conclusion} with a discussion of open problems and next steps. 
Note that throughout this work, unless otherwise specified we will restrict our attention to two-dimensional systems.

\section{Review: Twisted Boundary Conditions and the NTW Formula}
\label{TBC}
\begin{figure}
\includegraphics[width=\columnwidth]{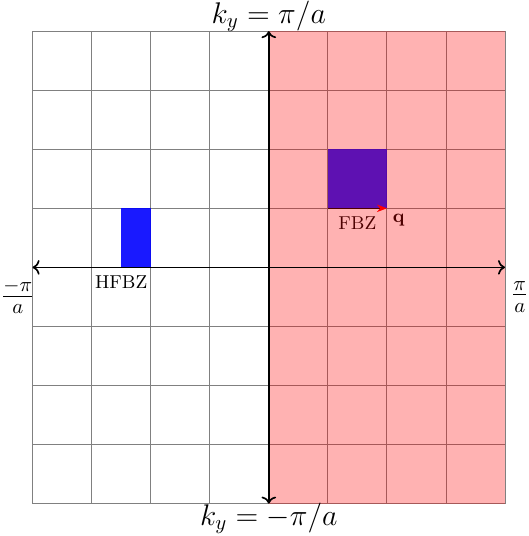}
% \begin{tikzpicture}

% \draw[step=1cm,color=gray] (-4,-4) grid (4,4);
%     \begin{scope}
%     	             ]
%         \fill[blue,fill opacity=.9] (1,1) rectangle (2,2);
%     \end{scope}
%     \begin{scope}
%     	             ]
%         \fill[blue,fill opacity=.9] (-3+0.5,0) rectangle (-2,1);
%     \end{scope}
%     \node at (1.5,0.75) {FBZ};
%     \node at (0.5-3,-0.25) {HFBZ};
%     \draw[-{Stealth[red]}]        (1,1)   -- (2,1);
%     \node at (2.2,0.8) {$\mathbf{q}$};
%     \pgfsetfillopacity{0.3}
%     \fill[red]   (0,-4) rectangle (4,4);
%     \draw[thick,<->] (-4,0) -- (4,0);
%     \draw[thick,<->] (0,-4) -- (0,4);
%     \pgfsetfillopacity{1}
%     \node at (4.25,-0.5) {\Large$\frac{\pi}{a}$};
%     \node at (-4.25,-0.5) {\Large$\frac{-\pi}{a}$};
%     \node at (0,4.25) {\Large$k_y=\pi/a$};
%     \node at (0,-4.25) {\Large$k_y=-\pi/a$};

% \end{tikzpicture}
\caption{Different regions within the Brillouin zone. $\mathbf{q}=2\pi(n_x/L_x,n_y/L_y)$ is a vector in the Brillouin zone (BZ). 
The red region is Half of the Brillouin zone (HBZ). 
The cross-shaded region (one of the squares in the Brillouin zone grid) is the Flux Brillouin zone (FBZ).}\label{BZs}
\end{figure}

In Ref. \cite{niuQuantizedHallConductance1985}, the authors (NTW) proposed a formula for the DC conductivity $\sigma_{xy}$ in terms of the Chern number of the ground state. 
They derive the Hall conductance of the ground state of a many-body system by using the Kubo formula, \cite{kubo_statistical-mechanical_1957}. 
They first obtain the current operator by twisting the boundary conditions on the ground state wavefunction (which is gauge-equivalent to adding a flux into the Hamiltonian, i.e., replacing $\mathbf{p}$ by $\mathbf{p}-\pmb{\alpha}$ for constant $\pmb{\alpha}$) and then differentiating the Hamiltonian with respect to the twist angles. 
They then use the Kubo formula to write the conductivity in terms of derivatives of the twisted Hamiltonian. 
Finally, they average the Hall conductance over all the twist angles to yield an integral of the ground state Berry curvature over the twist angles.  

We will now review the derivation of the NTW formula, as we will use it as a starting point for computing the many-body $\mathbb{Z}_2$ invariant.
We begin with a general a many-body Hamiltonian given by
\begin{equation}
    H=\sum_i\frac{1}{2m}[\mathbf{p}_i-e\mathbf{A}(\mathbf{r}_i)]^2+V.
\end{equation}
where, $i$ is a particle index, $\mathbf{A}$ is the vector potential corresponding to a potentially nonzero uniform external magnetic field $B\hat{z}=\nabla\cross\mathbf{A}$, and $V$ includes momentum-independent terms including single-particle potentials and two-body interactions.
The external magnetic field will not be essential for our purposes, but we retain it here for generality.
We introduce a constant flux parameter $\pmb{\alpha}$ in the kinetic term,
\begin{equation}\label{eq:general_ham_alpha}
    H\rightarrow H(\mathbf{\alpha})=\sum_i\frac{1}{2m}(\mathbf{p}_i-e\mathbf{A}(\mathbf{r}_i)+\hbar\pmb{\alpha})^2+V.
\end{equation}
Next, we look for the ground state of $H(\alpha)$ which satisfies (magnetic) periodic boundary conditions,
\begin{flalign}\label{eq:periodic_bcs}
    \mathcal{T}_{L_x}|\psi_0(\pmb{\alpha})\rangle=|\psi_0\rangle \nonumber\\
    \mathcal{T}_{L_y}|\psi_0(\pmb{\alpha})\rangle=|\psi_0\rangle.
\end{flalign}
The $\mathcal{T}_{L_a}$'s are the (magnetic) translation operators in a system of size $L_x\times L_y$,
\begin{equation}
    \mathcal{T}_{L_a}=\exp\left({i}\frac{L_aP_a}{\hbar}\right),
\end{equation}
where $P_a = \sum_i p_a^i-eA_a(\mathbf{r}_i) +(\mathbf{B}\cross \mathbf{r}_i)_a$ are the magnetic translation generators. 
Note that the system with Hamiltonian $H(\pmb{\alpha})$ with periodic boundary conditions \eqref{eq:periodic_bcs} is gauge equivalent to a system with Hamiltonian $\tilde{H}=H(\mathbf{0})$ with twisted boundary conditions $\mathcal{T}_{L_a}\ket{\tilde\psi} = e^{i\alpha_aL_a}\ket{\tilde\psi}$. 
This shows that $\alpha$ can be restricted to $[-\frac{2\pi}{L_x},\frac{2\pi}{L_x}]\times[-\frac{2\pi}{L_y},\frac{2\pi}{L_y}]$. 
This domain will be called the flux Brillouin zone (FBZ), and is shown in Fig.~\ref{BZs} as a subregion of the Brillouin zone associated to the crystal. 
We explain in appendix \ref{app:A} that shifting $\pmb{\alpha}$ by a Brillouin zone vector $\mathbf{q}=2\pi(n_x/L_x,n_y/L_y)$ is equivalent to a unitary transformation 
\begin{equation}\label{eq:alphaplusq}
|\psi_0(\pmb{\alpha}+\mathbf{q})\rangle={e}^{-{i}\mathbf{q}\cdot\mathbf{R}}|\psi_0(\pmb{\alpha})\rangle,
\end{equation} 
where $\mathbf{R}$ is the center of position $\mathbf{R}=\sum_i\mathbf{r}_i$. 
Thus the unitary transformation Eq.~\eqref{eq:alphaplusq} preserves the boundary conditions Eq.~\eqref{eq:periodic_bcs}.
In particular, we see from Eq.~\eqref{eq:alphaplusq}, we see that $\ket{\psi_0(\pmb{\alpha})}$ and $\ket{\psi_0(\pmb{\alpha}+\mathbf{q})}$ satisfy the same boundary conditions.
Moreover, the transformation $\ket{\psi(\pmb{\alpha})}\rightarrow e^{-i\mathbf{q}\cdot\mathbf{R}}\ket{\psi(\pmb{\alpha})}$ in Eq.~\eqref{eq:alphaplusq} is a (large) gauge transformation, and $H(\pmb{\alpha}+\mathbf{q})$ has the same spectrum as $H(\pmb{\alpha})$. 
Thus as long as $H(\pmb{\alpha})$ has a gapped ground state for all $\pmb{\alpha}$, adiabatic perturbation theory requires that $\ket{\psi_0(\pmb{\alpha})}$ and $\ket{\psi_0(\pmb{\alpha}+\mathbf{q})}$  be the same state (up to a phase) in the thermodynamic limit~\cite{resta1998quantum,resta2002insulators,bradlyn2022lecture}.

Once we have the many-body ground state as a function of $\pmb{\alpha}$, we compute the averaged DC conductivity by using the Kubo formula,
\begin{equation}
    \label{eq:24}
    \Bar{\sigma}_{xy}=-\frac{e^2}{2\pi{i}h}\int_{\text{FBZ}}\braket{\frac{\partial\psi_0}{\partial\alpha_x}}{\frac{\partial\psi_0}{\partial\alpha_y}}-\braket{\frac{\partial\psi_0}{\partial\alpha_y}}{\frac{\partial\psi_0}{\partial\alpha_x}}.
\end{equation}
The integrand is the Berry curvature on the flux torus. 
For noninteracting systems, it can be rewritten in terms of the momentum-space Berry curvature of the occupied Bloch states. 
In this way, Eq.~\eqref{eq:24} reduces to the well-known TKNN formula \cite{thoulessQuantizedHallConductance1982} (also see Appendix \ref{app:A}),
\begin{equation}
\label{cond}
    \sigma_{xy}=-\frac{e^2}{2\pi{i}h}\int_{\text{BZ}}\sum_{n=1}^N\braket{\frac{\partial u_n}{\partial k_x}}{\frac{\partial u_n}{\partial k_y}}-\braket{\frac{\partial u_n}{\partial k_y}}{\frac{\partial u_n}{\partial k_x}},
\end{equation}
where $|u_n(\mathbf{k})\rangle$ are the filled Bloch states for a noninteracting insulator.

We would like to extend this idea to a formula for the $\mathbb{Z}_2$ invariant. 
This invariant was first defined for non-interacting time-reversal invariant systems in Ref. \cite{c.l.kaneZ_2TopologicalOrder}. 
It was formulated as the number of zeroes of the Pfaffian of the TR operator in half of the Brillouin zone,
\begin{equation}
    P(\mathbf{k})=\text{Pf}\langle u_n(\mathbf{k})|\Theta|u_m(\mathbf{k})\rangle,
\end{equation}
modulo 2. 
This version of the $\mathbb{Z}_2$ invariant is difficult to generalize to many-body systems, as it relies on the Kramers doublet structure of the single-particle wavefunctions that is invisible to the many-body ground state. 
This is in stark contrast to the situation for the Chern number, where the many-body NTW formula Eq.~\eqref{eq:24} takes the same functional form as the TKNN formula Eq.~\eqref{cond}, when we replace $\mathbf{k}$ by $\pmb{\alpha}$ and the single-particle Bloch states $\ket{u_{n}(\mathbf{k})}$ with the many-body states $\ket{\psi_0(\pmb{\alpha})}$.
This motivates us to try to formulate the many-body $\mathbb{Z}_2$ invariant in terms of a Berry connection and Berry curvature.

\section{The Many-Body $\mathbb{Z}_2$ Invariant}
\label{many-body z2}

To build a formulation of the $\mathbb{Z}_2$ invariant in terms of a many-body Berry curvature, we will take inspiration from Ref. \cite{leeManyBodyGeneralizationTopological2008}, where a many-body formulation of the $\mathbb{Z}_2$ was provided. 
We consider a lattice with $N=L_x\times L_y=$odd number of lattice sites. 
Having an odd number of lattice sites is essential to guarantee that the Berry connection that we will introduce has the right transformation properties under time-reversal symmetry. 
The formulation of the invariant in Ref.~\cite{leeManyBodyGeneralizationTopological2008} is given in terms of the sewing matrix for time reversal $\Theta$ at high-symmetry points in the FBZ, rather than in terms of a Berry curvature. 
Nevertheless, we now show we can rewrite the results of Ref.~\cite{leeManyBodyGeneralizationTopological2008} in an equivalent way that makes the analogy to Eq.~\eqref{eq:24} clear. 

Let $|\psi(\pmb{\alpha})\rangle$ be the many-body ground state of a Hamiltonian $H(\pmb{\alpha})$ as in Eq.~\eqref{eq:general_ham_alpha}. 
We first define a set of quasi-single particle (QSP) states,
\begin{equation}
    \label{many-body spin states}    |A,\sigma(\pmb{\alpha})\rangle\propto\left(\prod_{\mathbf{k}\in\text{BZ}}c_{\mathbf{k}A\Bar{\sigma}}\right)|\psi(\pmb{\alpha})\rangle,
\end{equation}
where $A$ is a bipartite, momentum-independent, time-reversal even degree of freedom, and $\bar{\sigma}$ denotes the spin opposite to $\sigma$. 
For our models, we will choose $A$ to be a sublattice degree of freedom for concreteness. 
Care must be taken to both normalize and orthogonalize the QSPs in Eq.~\eqref{many-body spin states}. 
Note that no component of spin need be conserved to define the QSP states in Eq.~\eqref{many-body spin states}, but it is important that the QSP states do not vanish and are linearly independent. 
Ref. \cite{leeManyBodyGeneralizationTopological2008} defined analogous states by creating holes in position space rather than momentum space; since the HK-type ground states we will examine later are defined in momentum space, we choose to define Eq.~\eqref{many-body spin states} accordingly. 
This prescription differs from that of \cite{leeManyBodyGeneralizationTopological2008} by a Slater determinant which cancels out when we normalize the states in \eqref{many-body spin states}.
We fix a gauge such that
\begin{equation}
\label{TRstate1}
    \Theta|A,\uparrow(\pmb{\alpha})\rangle=|A,\downarrow(-\pmb{\alpha})\rangle
\end{equation}
\begin{equation}
\label{TRstate2}
    \Theta|A,\downarrow(\pmb{\alpha})\rangle=-|A,\uparrow(-\pmb{\alpha})\rangle.
\end{equation}
It is here that it becomes important that we consider $N$ odd, to ensure that the QSP states $\Theta\ket{A,\downarrow(\pmb{\alpha})}$ acquire the fermionic minus sign in Eq.~\eqref{TRstate2}. 
Finally, we define the $\mathbb{Z}_2$ invariant (for this many-body ground state) as,
\begin{equation}
\label{Z2_inv}
    \nu=\frac{1}{2\pi}\left[\oint_{\partial\text{HFBZ}}\!\!\!\!\!\!\!\!\mathrm{d}\pmb{\alpha}\cdot\mathbf{A}(\pmb{\alpha})-\int_{\text{HFBZ}}\!\!\!\!\!\!\!\!\mathrm{d}^2\pmb{\alpha}F_{xy}(\pmb{\alpha})\right]\!\!\!\!\mod 2,
\end{equation}
where 
\begin{equation}
\mathbf{A}(\pmb{\alpha})=\sum_{\sigma=\uparrow,\downarrow}\langle A,\sigma(\pmb{\alpha})|{i}\nabla|A,\sigma(\pmb{\alpha})\rangle
\end{equation}
is the (trace of the) Berry connection associated to the QSP states, and 
\begin{equation}
F_{xy}=\nabla\times\mathbf{A}
\end{equation}
is the associated (abelian) Berry curvature. 
The integrals in Eq.~\eqref{Z2_inv} are over half the flux Brillouin zone, denoted by HFBZ and shown in Fig.~\ref{BZs}. 
This is one of our main results, and is not contained in Ref. \cite{leeManyBodyGeneralizationTopological2008}. 

To show that $\nu$ is a well-defined topological invariant, we must show that it is quantized and gauge invariant. 
Quantization follows from Stokes's theorem, which guarantees that the integral in Eq.~\eqref{Z2_inv} is an integer. 
Furthermore, we argue that any gauge transformation consistent with the gauge constraints \eqref{TRstate1} and \eqref{TRstate2} can change $\nu$ only by an even integer, such that $\nu \mod 2$ is gauge invariant. 
To see this note that the flux Brillouin zone is compact. 
The gauge constraint fixes admissible gauge transformations of the QSP wavefunctions to be of the form
\begin{flalign}\label{eq:gaugechoice}
    &|A,\uparrow(\pmb{\alpha})\rangle\rightarrow{e}^{{i}\theta(\pmb{\alpha})}|A,\uparrow(\pmb{\alpha})\rangle\nonumber\\
    &|A,\downarrow(\pmb{\alpha})\rangle\rightarrow{e}^{-{i}\theta(\pmb{-\alpha})}|A,\downarrow(\pmb{\alpha})\rangle.
\end{flalign}
Under such a gauge transformation, $\nu$ in Eq. \eqref{Z2_inv} changes by
\begin{align}
\label{nuchange}
    \nu&\rightarrow\nu+\delta\nu \nonumber \\
    &\rightarrow\nu+\frac{1}{2\pi}\oint_{\partial\text{HFBZ}}\mathrm{d}\pmb{\alpha}\cdot(\nabla\theta(\pmb{\alpha})-\nabla\theta(\pmb{-\alpha})) \mod 2.
\end{align}
Thus, the change in $\nu$ is given by $\delta\nu = n_+-n_-$, the difference between the winding number $n_+$ of $\theta(\pmb{\alpha})$ and the winding number $n_-$ $\theta(\pmb{-\alpha})$ around the HFBZ. 
However, because the FBZ is compact, we must have that $n_+=-n_-$, since the winding of $\theta$ over the FBZ must vanish. 
Therefore, $\delta\nu=2n_+\in 2\mathbb{Z}$ is an even integer, and $\delta\nu\mod 2 =0$. 
Thus $\nu$ as defined in Eq.~\eqref{Z2_inv} is invariant under gauge transformations. 

Equation \eqref{Z2_inv} is equivalent to the formulation of the $\mathbb{Z}_2$ invariant in Ref. \cite{leeManyBodyGeneralizationTopological2008} as we show in Appendix \ref{app:B}. 
Our formula Eq. \eqref{Z2_inv} was inspired by the analogous formula for non-interacting systems in Ref. \cite{fuTimeReversalPolarization2006} given in terms of the Berry connections of the filled Bloch states. 
As we show in Appendix \ref{app:E}, the non-interacting formula follows from Eq.~\eqref{Z2_inv} by noting that $|\psi(\pmb{\alpha})\rangle=|\psi_{\uparrow}(\pmb{\alpha})\rangle|\psi_{\downarrow}(\pmb{\alpha})\rangle$ for non-interacting systems. 
Then, repeating the steps in the derivation of the TKNN formula from the NTW formula (see Appendix~\ref{app:A}) converts the integrals over the flux-Brillouin zone to integrals over the full Brillouin zone, yielding the non-interacting expression.

Building on this approach, in \ref{app:D}, we use a formula similar to \eqref{eq:24} and \eqref{Z2_inv} to calculate the electric polarization (mod $e$) of one-dimensional many-body insulators, as first outlined in~\cite{souza2000polarization} .
We apply their formula to the one-dimensional Su–Schrieffer–Heeger (SSH) model with an orbital HK interaction. 
Since the result of the calculation turns out to be $0 \mod e$, we define spin-resolved quasi-single-particle states, similar to the QSPs \eqref{TRstate1} and \eqref{TRstate2}. 
The many-body polarization of these states is $e/2$, consistent with the expected half polarization per spin in the SSH model.
Thus, the method outlined in this section can also be used to probe the topological structure of many-body ground states in one dimensional systems.

Finally, we note that there has been some progress in developing many-body formulations of the Kane-Mele and Fu-Kane formulas for the $\mathbb{Z}_2$ invariant in terms of single particle Green's functions. Some examples include references \cite{wang2010topologicalorder}, and \cite{qi2008topological}. However, as pointed out in Refs.\cite{zhao_failure_2023} and \cite{peralta2023connecting}, topological invariants defined in terms of Green's functions (such as $N_3$ in case of a Chern insulator) often fail to provide a correct classification of many-body systems when strong interactions are involved. In particular, the formation of zeroes in the single-particle Green's function (Luttinger surfaces) can herald the breakdown of correspondence between the winding of single-particle Green's functions and topological invariants. Nevertheless, we expect that in the limit of weak interactions (below the Mott scale), single-particle Green's function formulas should be applicable to the Hubbard model and will necessarily coincide with our Eqs.~\eqref{eq:24} and \eqref{Z2_inv}.

\section{Application to Hatsugai-Kohmoto Models} \label{HK}

We will now apply Eq.~\eqref{Z2_inv} to compute the $\mathbb{Z}_2$ invariant for a family of interacting systems with (orbital) Hatsugai-Kohmoto interactions, which are diagonal in momentum space. 
As a warmup in Sec.~\ref{sec:hkhaldane_main} we first show how the NTW formula Eq.~\eqref{eq:24} can be used to directly compute the Chern number for orbital Hatsugai Kohmoto models. 
Using lessons learned from this calculation, we will then proceed in Sec.~\ref{sec:kmhk_main} to calculate the $\mathbb{Z}_2$ invariant for the interacting Kane-Mele model with HK interaction.

\subsection{Chern numbers in the Haldane-HK Model}\label{sec:hkhaldane_main}
As shown in appendix \ref{app:A}, computing the conductivity for a non-interacting system using the NTW formula is equivalent to computing the Chern number of the Bloch states using the formula of TKNN~\cite{thoulessQuantizedHallConductance1982}. 
We will now apply the NTW formula to a square lattice model of a Chern insulator with orbital HK interactions (the Haldane-HK model). 
Our starting point is the Hamiltonian 
\begin{align}
 \label{HKham}
 H=\sum_{\mathbf{k}\in\text{BZ}}\Big[\sum_{\mu\mu'\sigma}c^{\dagger}_{\mathbf{k}\mu\sigma}[h(\mathbf{k})]_{\mu\mu'}c_{\mathbf{k}\mu'\sigma}+U\sum_{\mu}n_{\mathbf{k}\mu\uparrow}n_{\mathbf{k}\mu\downarrow}\Big],
\end{align}
where $[h(\mathbf{k})]_{\mu\mu'}$ is a single-body, spin-independent Bloch Hamiltonian. 
Greek indices $\mu,\mu'$ index the two basis orbitals $A$ and $B$ within the unit cell. 
For simplicity, we take $\mathrm{tr} [h(\mathbf{k})]=0$ We insert flux into this model by replacing $c_{\mathbf{x}\mu\sigma}$ with $\exp({i}\int^{\mathbf{x}}_0\mathbf{A})c_{\mathbf{x}\mu\sigma}={e}^{{i}\pmb{\alpha}\cdot\mathbf{x}}c_{\mathbf{x}\mu\sigma}$, which is equivalent to replacing the Bloch momentum $\mathbf{k}$ by $\mathbf{k}-\pmb{\alpha}$ in the non-interacting Hamiltonian. 
For constant $\pmb{\alpha}$, this gauge transformation leaves the HK interaction Hamiltonian unchanged~\cite{ma2024charge}. 
We thus find that the Hamiltonian as a function of $\pmb{\alpha}$ is given by
\begin{align}
\label{HKfluxham}    
   H(\pmb{\alpha})&=\sum_{\mathbf{k}\in\text{BZ}}\sum_{\mu\mu'\sigma}c^{\dagger}_{\mathbf{k}\mu\sigma}[h(\mathbf{k}-\pmb{\alpha})]_{\mu\mu'}c_{\mathbf{k}\mu'\sigma} \nonumber \\
 &+U\sum_{\mathbf{k}\in\text{BZ}}\sum_{\mu}n_{\mathbf{k}\mu\uparrow}n_{\mathbf{k}\mu\downarrow}.
\end{align}
We will focus on the two-particle ground state of this Hamiltonian. 
There are six basis states in the two-particle sector, which we give explicitly in Table.~\ref{basistable}. 

\begin{table}
\begin{tabular}{ ||c|c|c||} 
\hline
Index & State & Label \\
\hline

1& $c^{\dagger}_{\mathbf{k}A\uparrow}c^{\dagger}_{\mathbf{k}B\uparrow} |0\rangle$ & $|A\uparrow,B\uparrow\rangle$ \\ 
2& $c^{\dagger}_{\mathbf{k}A\uparrow}c^{\dagger}_{\mathbf{k}A\downarrow} |0\rangle$ & $|A\uparrow,A\downarrow\rangle$ \\
3& $c^{\dagger}_{\mathbf{k}B\uparrow}c^{\dagger}_{\mathbf{k}A\downarrow} |0\rangle$ & $|B\uparrow,A\downarrow\rangle$ \\
4& $c^{\dagger}_{\mathbf{k}A\uparrow}c^{\dagger}_{\mathbf{k}B\downarrow} |0\rangle$ & $|A\uparrow,B\downarrow\rangle$\\
5& $c^{\dagger}_{\mathbf{k}B\uparrow}c^{\dagger}_{\mathbf{k}B\downarrow} |0\rangle$ & $|B\uparrow,B\downarrow\rangle$\\
6& $c^{\dagger}_{\mathbf{k}A\downarrow}c^{\dagger}_{\mathbf{k}B\downarrow} |0\rangle$ & $|A\downarrow,B\downarrow\rangle$\\
\hline
\end{tabular}
\caption{The 6 basis states in the two-particle sector.}\label{basistable}
\end{table}

We can write down the matrix elements of the Hamiltonian \eqref{HKfluxham} in the two-particle basis, which gives us a $6\times 6$ matrix, $\mathcal{H}_g(\mathbf{k}-\pmb{\alpha})$. 
Next, note that states $1$ and $6$ in Table~\ref{basistable} consist of two electrons with the same spin in each unit cell; by the Pauli exclusion principle, the single-particle Hamiltonian annihilates basis states $1$ and $6$. 
Similarly, since these states have no double occupations, they are also annihilated by the HK interaction. 
Thus the only nontrivial eigenvectors of $\mathcal{H}_g(\mathbf{k}-\pmb{\alpha})$ come from linear combinations of basis states $2$--$5$. 
Diagonalizing $\mathcal{H}_g(\mathbf{k}-\pmb{\alpha})$ restricted to this nontrivial four-dimensional subspace gives us a lowest-energy eigenvector $U^{\mathrm{g.s.}}_{\mu\mu'}(\mathbf{k}-\pmb{\alpha})$,
from which we write down the ground state of \eqref{HKfluxham} as
\begin{align}
    \label{hkgsflux}
    |\psi_0(\pmb{\alpha})\rangle&=\prod_{\mathbf{k}\in\text{BZ}}\left(\sum_{\mu\mu'}U^{\text{g.s.}}_{\mu\mu'}(\mathbf{k}-\pmb{\alpha})c^{\dagger}_{\mathbf{k}\mu\uparrow}c^{\dagger}_{\mathbf{k}\mu'\downarrow}\right)|0\rangle\nonumber\\&=\prod_{\mathbf{k}\in\text{BZ}}|\psi_{\mathbf{k}}(\pmb{\alpha})\rangle,
\end{align}
where we have exploited the fact that all basis states $2$--$5$ have antialigned spins, allowing us to suppress the spin index on $U^{\mathrm{g.s.}}_{\mu\mu'}(\mathbf{k}-\pmb{\alpha})$. 
Plugging Eq.~\eqref{hkgsflux} into \eqref{eq:24}, we get,
\begin{widetext}
\begin{flalign}
    \bar{\sigma}_{xy}&=-\frac{e^2}{h}\frac{1}{2\pi{i}}\sum_{\mathbf{k}\in\text{BZ}}\int\mathrm{d}^2\pmb{\alpha}\sum_{\mu\mu'}\left(\frac{\partial U^*_{\mu\mu'}(\mathbf{k}-\pmb{\alpha})}{\partial\alpha_x}\frac{\partial U_{\mu\mu'}(\mathbf{k}-\pmb{\alpha})}{\partial\alpha_y}-\frac{\partial U^*_{\mu\mu'}(\mathbf{k}-\pmb{\alpha})}{\partial\alpha_y}\frac{\partial U_{\mu\mu'}(\mathbf{k}-\pmb{\alpha})}{\partial\alpha_x}\right)\\
    &=-\frac{e^2}{h}\frac{1}{2\pi{i}}\int_{\text{CBZ}}\mathrm{d}^2\mathbf{k}\sum_{\mu\mu'}\left(\frac{\partial U^*_{\mu\mu'}(\mathbf{k})}{\partial k_x}\frac{\partial U_{\mu\mu'}(\mathbf{k})}{\partial k_y}-\frac{\partial U^*_{\mu\mu'}(\mathbf{k})}{\partial k_y}\frac{\partial U_{\mu\mu'}(\mathbf{k})}{\partial k_x}\right),\label{eq:hkntw}
\end{flalign}
\end{widetext}
where in the second line, we replaced the sum over the discrete BZ momentum $\mathbf{k}$ and the integral over the continuous $\pmb{\alpha}$, with an integral over continuous $\mathbf{k}$ in the continuous Brillouin zone (CBZ) (the $\pmb{\alpha}$'s occupy the spaces between the $\mathbf{k}$'s, as shown in Fig.~\ref{BZs}). 
This shows that the Chern number of the ground state $\ket{\psi_0(\pmb{\alpha})}$ can be evaluated as the Chern number of $U_{\mu\mu'}^{\text{g.s.}}(\mathbf{k})$ in the thermodynamic limit.  We use this method to compute the Chern number of the Haldane-HK model in Appendix \ref{app:C}.

Previously,  Refs.~\cite{10.1093/acprof:oso/9780199564842.001.0001,PhysRevB.83.085426} proposed that the Chern number of an interacting system could be captured by an invariant $N_3$ defined in terms of the many-body Green's function. $N_3$ reduces to the TKNN invariant in the non-interacting limit and thus is a natural candidate for generalization of the Chern invariant to many-body systems. 
However, Refs.~\cite{zhao_failure_2023,peralta2023connecting} showed that $N_3$ deviates strongly from the Hall conductance in the Haldane-HK model, which contains zeroes in the Green's function. 
In particular, Ref.~\cite{zhao_failure_2023} computed the Chern number for the Haldane-HK model directly from the Kubo formula for the Hall conductivity. 
Here, we show alternatively that the Chern number can be computed directly from the NTW formula, and that these two calculations agree.

\subsection{$\mathbb{Z}_2$ Invariant in the Kane-Mele-HK model}\label{sec:kmhk_main}
\begin{figure}[h]
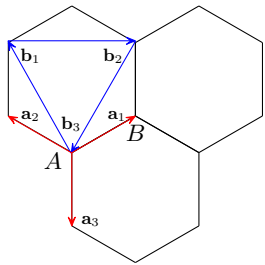


    \centering
    \setchemfig{bond style={black}}
    \chemfig{*6(-*6(------)-*6(------)----)}
    \chemmove{\draw [blue] (-1.8,1)--(-2.66,-0.495);
              \draw [blue] (-2.66,-0.495)--(-3.5,1);
              \draw [blue] (-3.5,1)--(-1.8,1);
              \draw [red] (-2.66,-0.495)--(-1.8,0);
              \draw [red] (-2.645,-0.495)--(-3.5,0);
              \draw [red] (-2.65,-0.495)--(-2.65,-1.47);
              \node [scale=0.75] at (-3.2,0.8) {$\mathbf{b}_1$};
              \node [scale=0.75] at (-2.66,-0.1) {$\mathbf{b}_3$};
              \node [scale=0.75] at (-2.1,0.8) {$\mathbf{b}_2$};
              \node [scale=0.75] at (-2.05,0) {$\mathbf{a}_1$};
              \node [scale=0.75] at (-3.2,0) {$\mathbf{a}_2$};
              \node [scale=0.75] at (-2.4,-1.4) {$\mathbf{a}_3$};
              \node at (-2.9, -0.6) {$A$};
              \node at (-1.8, -0.23) {$B$};}
    \caption{This is the graphene lattice we will be using, with $\mathbf{a}_1=(\frac{\sqrt{3}}{2},\frac{1}{2})$, $\mathbf{a}_2=(-\frac{\sqrt{3}}{2},\frac{1}{2})$ and $\mathbf{a}_3=(0,-1)$. 
    We also have the next-nearest neighbor lattice vectors, $\mathbf{b}_1=(-\frac{\sqrt{3}}{2},\frac{3}{2})$, $\mathbf{b}_2=(\sqrt{3},0)$ and $\mathbf{b}_3=(-\frac{\sqrt{3}}{2},-\frac{3}{2})$. }
    \label{fig:enter-label}

\end{figure}

Now that we have seen how to evaluate the Chern number via the NTW formula for the Haldane-HK model, we will apply these lessons to the The Kane-Mele model with an HK interaction. 
In particular, we will show how our formula \eqref{Z2_inv} for the $\mathbb{Z}_2$ invariant can be used to calculate $\nu$ for the Kane-Mele model with an HK interaction. 
Our starting point is the Hamiltonian $H=H_0+V$,
\begin{align}\label{eq:hkkmham}
  H_0&=t\sum_{\mathbf{k}\sigma}g(\mathbf{k})c^{\dagger}_{\mathbf{k}A\sigma}c_{\mathbf{k}B\sigma}+\text{h.c.}+\lambda_v\sum_{\mathbf{k}\sigma\mu}\text{sign}(\mu)c^{\dagger}_{\mathbf{k}\mu\sigma}c_{\mathbf{k}\mu\sigma}\notag\\&+2\lambda_1\sum_{\mathbf{k}\sigma\mu}\Tilde{g}_1(\mathbf{k})\text{sign}(\mu)\text{sign}(\sigma)c^{\dagger}_{\mathbf{k}\mu\sigma}c_{\mathbf{k}\mu\sigma}
\end{align}
and
\begin{equation}
    V=U\sum_{\mathbf{k}\mu}n_{\mathbf{k}\mu\uparrow}n_{\mathbf{k}\mu\downarrow},
\end{equation}
where,
\begin{equation}
    g(\mathbf{k})=\sum_{i=1}^3 {e}^{{i}\mathbf{k}\cdot\mathbf{a}_i},
\end{equation}
and 
\begin{equation}
    \Tilde{g}_1(\mathbf{k})=\sum_{i=1}^3\sin(\mathbf{k}\cdot\mathbf{b}_i).
\end{equation}
Here $\mu=A,B$ indexes the sublattice sites in the honeycomb lattice. 
The vectors $\mathbf{a}_i$ are the vectors between neighboring honeycomb lattice sites, and the $\mathbf{b}_i$ are the vectors to the next nearest neighbor sites. 
We show the $\mathbf{a}_i$'s and $\mathbf{b}_i$'s explicitly in Fig.~\ref{fig:enter-label}

Since the Hamiltonian \eqref{eq:hkkmham} is local in $\mathbf{k}$ space, we can diagonalize it by first writing $H=\sum_{\mathbf{k}}H(\mathbf{k})$ and then diagonalizing each $H(\mathbf{k})$ separately. 
We will focus on the two-particle sector, spanned by the six basis states in table \ref{basistable}. 
The Hamiltonian density in this basis is given by,
\begin{equation}
\label{eq:24.1}
H(\mathbf{k})=
\begin{pmatrix}
    0&0&0&0&0&0\\
    0&U+2\lambda_v&tg&tg&0&0\\
    0&tg^*&-4\lambda_1\Tilde{g}_1&0&tg&0\\
    0&tg^*&0&4\lambda_1\Tilde{g}_1&tg&0\\
    0&0&tg^*&tg^*&U-2\lambda_v&0\\
    0&0&0&0&0&0\\
\end{pmatrix}.
\end{equation}
Only the middle block of Hamiltonian \eqref{eq:24.1}, spanned by $\{c^{\dagger}_{\mathbf{k}A\uparrow}c^{\dagger}_{\mathbf{k}A\downarrow},c^{\dagger}_{\mathbf{k}B\uparrow}c^{\dagger}_{\mathbf{k}A\downarrow},c^{\dagger}_{\mathbf{k}A\uparrow}c^{\dagger}_{\mathbf{k}B\downarrow},c^{\dagger}_{\mathbf{k}B\uparrow}c^{\dagger}_{\mathbf{k}B\downarrow}\}$ acting on $|0\rangle$ is non trivial. 
The characteristic equation for the eigenvalues $x$ of this $4\times4$ block is,
\begin{align}
\label{eq:6}
    &(U-x)(U-x)(x^2-16\lambda_1^2\Tilde{g}_1^2)\\\nonumber&+4|tg|^2x(U-x)-4\lambda_v^2(x^2-16\lambda_1^2\Tilde{g}_1^2)=0.
\end{align}
In order to compute the $\mathbb{Z}_2$ invariant, we will look at the ground state of \eqref{eq:24.1} in two limits, $\lambda_1=0$ and $\lambda_v=0$, then use results from topology to argue that there is a phase transition as we vary the three couplings $\lambda_1,\lambda_v$ and $U$.
\subsubsection{$\lambda_1=0$}
\label{section 2}
In this regime, the middle block of \eqref{eq:24.1} is given by,
\begin{equation}
\label{eq:26.1}
    H(\mathbf{k})=
\begin{pmatrix}
    U+2\lambda_v&tg&tg&0\\
    tg^*&0&0&tg\\
    tg^*&0&0&tg\\
    0&tg^*&tg^*&U-2\lambda_v\\
\end{pmatrix}.
\end{equation}
The characteristic equation \eqref{eq:6} reduces to
\begin{equation}
\label{eq:27.1}
    (U-x)x\left[(U-x)x+4|tg|^2\right]=4\lambda_v^2x^2.
\end{equation}

From the characteristic equation \eqref{eq:27.1}, we see that $x=0$ is an eigenvalue at every $\mathbf{k}$ point. 
Therefore, the ground state energy must be negative everywhere. 
Also if we look at the $K$ point in the graphene Brillouin zone, where $g(K)=0$, we find that the three roots of \eqref{eq:27.1} are $x=0,U\pm2\lambda_v$. 
Thus, to avoid band-crossings, $0<U<2\lambda_v$. 
With this constraint, we can look at the characteristic equation for the three nontrivial roots,
\begin{equation}
\label{eq:29.1}
    p(x)=(U-x)\left[(U-x)x+4|tg|^2\right]-4\lambda_v^2x=0.
\end{equation}
We now argue that the smallest root of equation \eqref{eq:29.1} has multiplicity one, i.e., there are no degeneracies for all choices of $|tg|$ and $0<U<2\lambda_v$. 
To see this, note that if the ground state has multiplicity two, then, from the coefficient of $x^2$ we require
\begin{equation}
\label{eq:10}
    2E_g+E_2=2U>0,
\end{equation}
and from the coefficient of $x^0$ we require
\begin{equation}
\label{eq:11}
    E_g^2E_2=-4U|tg|,
\end{equation}
where $E_g<0$ and $E_2$ are the two roots ($E_g$ has multiplicity two). 
It is impossible to satisfy these equations: from \eqref{eq:10}, $E_2$ must be positive since $E_g<0$ but then it will be impossible to satisfy \eqref{eq:11}. 
Hence, for all choices of $0<U<2\lambda_v$, the ground state energy is non-degenerate everywhere in the Brillouin zone. 
This allows us to tune $U$ to zero without changing the $\mathbb{Z}_2$ invariant, which we know is zero at $U=0$. 
Therefore, for $\lambda_1=0$, the system is always in the trivial phase as long as $0<U<2\lambda_v$, after which it becomes a metal. 

In fact, if we consider the full Hamiltonian \eqref{eq:24.1}, $H=H_0+V$, we see that it has a non-degenerate ground state and we can check for a phase transition by looking for a band inversion at the $K$ point. 
The characteristic polynomial at the $K$ point is,
\begin{equation}
    (x^2-16\lambda_1^2\Tilde{g}_1^2)[(U-x)^2-4\lambda_v^2]=0,
\end{equation}
which has four roots, $x=\{\pm4\lambda_1\Tilde{g}_1(K),U\pm2\lambda_v\}=\{\pm6\sqrt{3}\lambda_1,U\pm2\lambda_v\}$. 
The two lowest energy states are, $x=-6\sqrt{3}\lambda_1$ and $x=U-2\lambda_v$. 
When the former state is lowest in energy, we are in the non-trivial QSH phase, whereas if the latter state is the lowest energy state, we are in the trivial phase. 
This phase transition happens precisely when $6\sqrt{3}\lambda_1=2\lambda_v-U$ or when $3\sqrt{3}\lambda_1=\lambda_v-\frac{U}{2}$.

Now, we support this analysis by calculating the many-body $\mathbb{Z}_2$ invariant analytically for $0<U<2\lambda_v$. 
The ground state for \eqref{eq:26.1} is given by,
\begin{equation}
    |\psi_0(\pmb{\alpha})\rangle=\prod_{\mathbf{k}\in\text{BZ}}\left(\sum_{\mu\mu'}U^{\text{g.s.}}_{\mu\mu'}(\mathbf{k}-\pmb{\alpha})c^{\dagger}_{\mathbf{k}\mu\uparrow}c^{\dagger}_{\mathbf{k}\mu'\downarrow}\right)|0\rangle,
\end{equation}
where $U_{\mu\mu'}(\mathbf{k})$ is the eigenvector of Eq.~\eqref{eq:26.1} with eigenvalue $E_g$. 
It can be written as 
\begin{align}
\label{eq:9}
    U^{\mathrm{g.s.}}(\mathbf{k})=\frac{1}{\sqrt{N}_{\mathbf{k}}}
    \begin{pmatrix}
        2tg(E_g-U+2\lambda_v)\\
        (E_g-U)^2-4\lambda_v^2\\
        (E_g-U)^2-4\lambda_v^2\\
        2tg^*(E_g-U-2\lambda_v)
    \end{pmatrix},
\end{align}
where the normalization constant $N_{\mathbf{k}}$ is given by
\begin{equation}
N_{\mathbf{k}}=8|tg|^2((E_g-U)^2+4\lambda_v^2))+2((E_g-U)^2-4\lambda_v^2)^2.
\end{equation}
{One needs to check that \eqref{eq:9} is well defined at all points inside the Brillouin zone. 
At a distance $\epsilon$ from the $K$ or $K^\prime$ points in the graphene Brillouin zone, $|tg|\sim\mathcal{O}(\epsilon)$, $E_g-U+2\lambda_v\sim\mathcal{O}(\epsilon^2)$, $E_g-U-2\lambda_v\sim\mathcal{O}(1)$ and $\sqrt{N}_{\mathbf{k}}\sim\mathcal{O}(\epsilon)$. 
From this, we see that close to the $K$ and $K^\prime$ points, $U^{\mathrm{g.s.}}$ is neither divergent nor vanishing.}

From Eq.~\eqref{many-body spin states} we then find that single particle states (after normalization; at half filling the QSP states with fixed sublattice index are already orthogonal) are
\begin{widetext}
\begin{flalign}\label{eq:km1pstates}
&|A,\sigma(\pmb{\alpha})\rangle=\prod_{\mathbf{k}\in\text{BZ}}\frac{1}{\sqrt{M^-}}\left(2tgc^{\dagger}_{\mathbf{k}A\sigma}+(E_g-U-2\lambda_v)c^{\dagger}_{\mathbf{k}B\sigma}\right)|0\rangle\nonumber\\
&|B,\sigma(\pmb{\alpha})\rangle=\prod_{\mathbf{k}\in\text{BZ}}\frac{1}{\sqrt{M^+}}\left((E_g-U+2\lambda_v)c^{\dagger}_{\mathbf{k}A\sigma}+2tg^*c^{\dagger}_{\mathbf{k}B\sigma}\right)|0\rangle,
\end{flalign}
\end{widetext}
where 
\begin{equation}
M^{\pm}=4|tg|^2+(E_g-U\pm 2\lambda_v)^2.
\end{equation}
Importantly, note that the coefficients of the creation operators, and the normalization factor are evaluated at $\mathbf{k}-\pmb{\alpha}$. 
Additionally, within $\epsilon$ of the the $K$ or $K'$ point $M^-\sim\mathcal{O}(1)$ and $M^+\sim \mathcal{O}(\epsilon^2)$. 
Thus $|A,\sigma(\pmb{\alpha})\rangle$ is manifestly regular at $K$ and $K'$, while for $|B,\sigma(\pmb{\alpha})\rangle$ we note that the coefficients in Eq.~\eqref{eq:km1pstates} are of order $\epsilon$ and $\epsilon^2$ near $K$ and $K'$, so that $|B,\sigma(\pmb{\alpha})\rangle$ is regular as well. 
Thus the single-particle states are well-defined and normalized for all $\mathbf{k}$.

The states \eqref{eq:km1pstates} satisfy the gauge constraints \eqref{TRstate1} and \eqref{TRstate2} if there are an odd number of points in the Brillouin zone.
We can thus use Eq.~\eqref{Z2_inv} to evaluate the $\mathbb{Z}_2$ invariant. 
We can use the structure of the FBZ to replace the integrals and derivatives over the $\pmb{\alpha}$'s in Eq. \eqref{Z2_inv} with integrals and derivatives over the $\mathbf{k}$'s in the HBZ, as shown in Appendix~\ref{app:D}.
Then for the two sublattices $A$ and $B$, we can write 
\begin{equation}
\label{eq:49}
    \nu^{A/B}=\frac{1}{2\pi}\left[\oint_{\partial\text{HBZ}}\!\!\!\!\!\!\!\!\mathrm{d}\mathbf{k}\cdot\mathbf{A}^{A/B}(\mathbf{k})-\int_{\text{HBZ}}\!\!\!\!\!\!\!\!\mathrm{d}^2\mathbf{k}F^{A/B}_{xy}(\mathbf{k})\right]\!\!\!\!\!\!\! \mod 2
\end{equation}
where $\nu^{A/B}$ is the $\mathbb{Z}_2$ invariant computed using either the $\ket{A,\sigma(\pmb{\alpha})}$ or $\ket{B,\sigma(\alpha)}$ QSP states.  $\mathbf{A}^{A/B}=\mathbf{A}^{A/B}_{\uparrow}+\mathbf{A}^{A/B}_{\downarrow}$ is the corresponding Berry connection given by
\begin{equation}
\label{A+-}
    \mathbf{A}^{A/B}(\mathbf{k})=\frac{8|tg|^2}{4|tg|^2+(E_g-U\mp 2\lambda_v)^2}\nabla_{\mathbf{k}}\theta,
\end{equation}
where $tg=|tg|{e}^{{i}\theta}$, and the minus sign choice corresponds to the $A$ sublattice. 
These Berry connections have well-defined partial derivatives everywhere inside the half Brillouin zone except at the $K$ point. 
Applying Stokes's theorem to the surface integral in Eq.~\eqref{eq:49}, we see that the surface integral also gives a line integral of the Berry connection around the boundary of the half Brillouin zone. 
However, the gauge for this line integral is required to be one in which the Berry connection is well defined at every point inside the half Brillouin zone, including the $K$ point; this need not be the same gauge as Eq.~\eqref{A+-}, which is used to evaluate the first term of Eq.~\eqref{eq:49}. 
Thus, since all singularities of Eq.~\eqref{A+-} occur at the $K$ or $K'$ points (and the $K'$ point is outside the HBZ), the difference of integrals in \eqref{eq:49} reduces to the winding of the Berry connection \eqref{A+-} around small circle of radius $\epsilon$ around the $K$ point. 

From Eq.~\eqref{eq:29.1}, we see that within $\epsilon$ of the $K$ point, the ground state energy behaves like,
\begin{equation}
    E_g\sim U-2\lambda_v + \mathcal{O}(\epsilon^2).
\end{equation} 
Therefore, we have 
\begin{align}
\mathbf{A}^{B}(K+\epsilon\mathbf{k})&\sim 2\nabla\theta \nonumber \\
 \mathbf{A}^{A}(K+\epsilon\mathbf{k})&\propto\epsilon^2\nabla\theta.
\end{align} 
Since $\theta$ has winding $2\pi$ around the $K$ point, for the first connection, we get $\nu^B=2\equiv 0\mod 2$ and plugging in the second connection, we get $\nu^A=0$ when we take the $\epsilon\rightarrow 0$ limit. 
Thus, graphene with a sublattice splitting potential is a trivial insulator, even in the presence of the HK interaction, and the choice of sublattice used in defining the states in \eqref{many-body spin states} does not make a difference.
\subsubsection{$\lambda_v=0$}
\label{section 3}
Having seen how to evaluate the many-body $\mathbb{Z}_2$ invariant in the topologically trivial phase, we now turn to the more interesting case where $\lambda_v=0$, with nonzero spin-orbit coupling $\lambda_1$. 
In this regime, the middle block of the Hamiltonian \eqref{eq:24.1} takes the form
\begin{align}
\label{eq:12}
    H(\mathbf{k})=
    \begin{pmatrix}
    U&tg&tg&0\\
    tg^*&-4\lambda_1\Tilde{g}_1&0&tg\\
    tg^*&0&4\lambda_1\Tilde{g}_1&tg\\
    0&tg^*&tg^*&U\\
\end{pmatrix}.
\end{align}
The characteristic equation of \eqref{eq:12} is,
\begin{equation}
\label{eq:13}
    (U-x)\left[(U-x)(x^2-16\lambda_1^2\Tilde{g}_1^2)+4|tg|^2x\right]=0,
\end{equation}
which gives $x=U$ as a trivial solution. 
One can see that this equation also does not have a degenerate ground state. 
If there were a degeneracy, the nontrivial roots of Eq. \eqref{eq:13} will satisfy,
\begin{flalign}
    &2E_g+E_2=U\\
    &E_g^2E_2=-16U\lambda_1^2\tilde{g}_1^2<0.
\end{flalign}
From the second equation, we see that $E_2$ must be negative. 
Since $E_g\leq E_2$, $E_g<0$. 
But that would make the first equation impossible to solve. 
Hence, the ground state cannot be degenerate. 
The ground state of \eqref{eq:12} is,
\begin{equation}
\label{eq:14}
    U^{\mathrm{g.s.}}(\mathbf{k})=\frac{1}{\sqrt{N_\mathbf{k}}}
    \begin{pmatrix}
        (E_g^2-16\lambda_1^2\Tilde{g}_1^2)tg\\
        2|tg|^2(E_g-4\lambda_1\Tilde{g}_1)\\
        2|tg|^2(E_g+4\lambda_1\Tilde{g}_1)\\
        (E_g^2-16\lambda_1^2\Tilde{g}_1^2)tg^*\\
    \end{pmatrix},
\end{equation}
where $N_{\mathbf{k}}=2(E_g^2-16\lambda_1^2\Tilde{g}_1^2)^2|tg|^2+8|tg|^4(E_g^2+16\lambda_1^2\Tilde{g}_1^2)$ is the normalization constant \footnote{One can check that $|\psi_{\mathbf{k}}\rangle$ is well defined near the $K$ point since $\sqrt{N}_{\mathbf{k}}\sim\epsilon^2$, but each component in the numerator scales at least as $\epsilon^2$.}.
\begin{widetext}
The normalized QSPs corresponding to the ground state are (again writing $tg=|tg|{e}^{{i}\theta}$), 
\begin{flalign}\label{eq:km_lambda_1_Astates}
    &|A,\uparrow(\pmb{\alpha})\rangle=\prod_{\mathbf{k}\in\text{BZ}}\frac{1}{\sqrt{M_+}}\left((E_g+4\lambda_1\Tilde{g}_1){e}^{{i}\theta}c^{\dagger}_{\mathbf{k}A\uparrow}+2|tg|c^{\dagger}_{\mathbf{k}B\uparrow}\right)|0\rangle\nonumber\\
    &|A,\downarrow(\pmb{\alpha})\rangle=\prod_{\mathbf{k}\in\text{BZ}}\frac{1}{\sqrt{M_-}}\left((E_g-4\lambda_1\Tilde{g}_1){e}^{{i}\theta}c^{\dagger}_{\mathbf{k}A\downarrow}+2|tg|c^{\dagger}_{\mathbf{k}B\downarrow}\right)|0\rangle,
\end{flalign}
where $M_{\pm}=(E_g\pm4\lambda_1\Tilde{g}_1)^2+4|tg|^2$. 
They satisfy the gauge conditions. 
If we choose the B sublattice, we get the following states,
\begin{flalign}
    &|B,\uparrow(\pmb{\alpha})\rangle=\prod_{\mathbf{k}\in\text{BZ}}\frac{1}{\sqrt{M_-}}\left(2|tg|c^{\dagger}_{\mathbf{k}A\uparrow}+(E_g-4\lambda_1\Tilde{g}_1){e}^{-{i}\theta}c^{\dagger}_{\mathbf{k}B\uparrow}\right)|0\rangle\nonumber\\
    &|B,\downarrow(\pmb{\alpha})\rangle=\prod_{\mathbf{k}\in\text{BZ}}\frac{1}{\sqrt{M_+}}\left(2|tg|c^{\dagger}_{\mathbf{k}A\downarrow}+(E_g+4\lambda_1\Tilde{g}_1){e}^{-{i}\theta}c^{\dagger}_{\mathbf{k}B\downarrow}\right)|0\rangle.
\end{flalign}
\end{widetext}
From here on, we will focus on the A sublattice states without loss of generality. 
The B sublattice states lead to the same Berry connection as the A sublattice states, up to a minus sign, and thus to the same value of $\nu$.  
The Berry connection for the QSP states in Eq.~\eqref{eq:km_lambda_1_Astates} is given by
\begin{equation}\label{eq:A_nontrivial}
    \mathbf{A}(\mathbf{k})=\left[\frac{(E_g-4\lambda_1\Tilde{g}_1)^2}{M^-_{\mathbf{k}}}+\frac{(E_g+4\lambda_1\Tilde{g}_1)^2}{M^+_{\mathbf{k}}}\right]\nabla_{\mathbf{k}}\theta.
\end{equation}
We again need to evaluate this connection near the $K$ point as that is the only point in the HBZ where $\nabla_\mathbf{k}\theta$ has singularities. 
As in Sec.~\ref{section 2}, the difference of integrals in Eq.~\eqref{eq:49} for $\nu$ can be reduced, in this gauge, to the winding of Eq.~\eqref{eq:A_nontrivial} about the singularity at $K$. 
Since $E_g\sim-4\lambda_1\Tilde{g}_1$ near the $K$ point, we have $(E_g+4\lambda_1\Tilde{g}_1)^2\sim\mathcal{O}(\epsilon^4)$. 
Therefore,
\begin{equation}
    \mathbf{A}(K+\epsilon\mathbf{k})=(1+\mathcal{O}(\epsilon^2))\nabla_{\mathbf{k}}\theta,
\end{equation}
where the leading order term comes from the first term in Eq.~\eqref{eq:A_nontrivial}. 
Since $\theta$ has a winding of $2\pi$ around the $K$ point, we find $\nu=1$. 
Thus, we have shown that the Kane-Mele model in the presence of the HK interaction remains nontrivial for $U>0$ when $\lambda_1\neq 0, \lambda_v=0$.

\subsubsection{Many-body spin Chern number}
An alternative formulation of the $\mathbb{Z}_2$ invariant that is sometimes used for systems with (approximate) spin-conservation is in terms of the spin-Chern number of the ground state. 
As originally argued in Ref.~\cite{sheng2006quantum}, the $\mathbb{Z}_2$ invariant is given by the parity of the Chern number evaluated for a system with spin-dependent twisted boundary conditions, evaluated using the NTW formula. 
For noninteracting systems, the spin-Chern number can also be defined for systems without conserved spin~\cite{prodan2009robustness,lin2024spin}. 
Within our framework, we can apply these ideas by computing the spin Chern number via the QSP states using the results of Sec.~\ref{sec:hkhaldane_main}. 
As argued in Ref.~\cite{leeManyBodyGeneralizationTopological2008}, this gives an equivalent expression for the $\mathbb{Z}_2$ invariant. 
For the Kane-Mele-HK model, we will now compute the spin Chern numbers as the Chern numbers of the $|A,\uparrow(\pmb{\alpha})\rangle$ states via the generalized NTW formula.

First, we compute the Chern number of $|A,\uparrow(\pmb{\alpha})\rangle$ in the nontrivial case where $\lambda_v=0$. 
This is equivalent to computing the Chern number of the two component state
\begin{equation}
\label{psiup}
    |\psi_{\uparrow}(\mathbf{k})\rangle=\frac{1}{\sqrt{M^+_{\mathbf{k}}}}\begin{pmatrix}
        (E_g+4\lambda_1\tilde{g}_1)\\
        2tg^*
    \end{pmatrix}
\end{equation}
defined from the amplitudes in Eq.~\eqref{eq:km_lambda_1_Astates}, where $E_g$ is the smallest eigenvalue of Eq. \eqref{eq:12} and $M^+_{\mathbf{k}}=(E_g+4\lambda_1\Tilde{g}_1)^2+4|tg|^2$. 
We will compute the Chern number numerically using the Mathematica code in provided at Ref.~\cite{cherncode}, which implements the surface integral of the Berry curvature in terms of a sum of discrete Wilson loops as first specified in Ref.~\cite{doi:10.1143/JPSJ.74.1674}. 
This code finds the Chern number of the lowest band of any $2\times 2$ Bloch Hamiltonian on a discretized Brillouin zone. 
We construct a flat band Hamiltonian whose negative energy eigenfunction is given by Eq. \eqref{psiup} by first finding a state orthogonal to $|\psi_{\uparrow}(\mathbf{k})\rangle$,
\begin{equation}
    |\psi_{\downarrow}(\mathbf{k})\rangle=\frac{1}{\sqrt{M^+_{\mathbf{k}}}}\begin{pmatrix}
        2tg\\
        -(E_g+4\lambda_1\tilde{g}_1)
    \end{pmatrix}
\end{equation}
and then constructing the 2×2 Bloch Hamiltonian,
\begin{equation}
    H(\mathbf{k})=|\psi_{\downarrow}(\mathbf{k})\rangle\langle\psi_{\downarrow}(\mathbf{k})|-|\psi_{\uparrow}(\mathbf{k})\rangle\langle\psi_{\uparrow}(\mathbf{k})|.
\end{equation}
Computing the Chern number $\nu_\uparrow$ associated to the negative energy state of this Hamiltonian, we find $\nu_{\uparrow}=1$. 
Thus we verify, as argued in Ref.~\cite{leeManyBodyGeneralizationTopological2008}, that $-1=(-1)^\nu = (-1)^{\nu_\uparrow}$ for the nontrivial phase of the Kane-Mele model with HK interactions. 

We can repeat the same analysis for the trivial phase $\lambda_1=0,\lambda_v\neq 0$ which we considered in section \ref{section 2}. 
From Eq.~\eqref{eq:km1pstates} we can write the $\ket{A,\uparrow(\mathbf{k})}$ states as a two component vector
\begin{equation}
\label{spinchernnumberstatefortrivial}
    |\phi_{\uparrow}(\mathbf{k})\rangle=\frac{1}{\sqrt{M^-_{\mathbf{k}}}}\begin{pmatrix}
        2tg\\
        E_g-U-2\lambda_v
    \end{pmatrix}.
\end{equation}
where $E_g$ is now the smallest eigenvalue of Eq. \eqref{eq:26.1} and $M^-_{\mathbf{k}}=\sqrt{4|tg|^2+(E_g-U-2\lambda_v)^2}$. 
The state orthogonal to Eq. \eqref{spinchernnumberstatefortrivial} is,
\begin{equation}
    |\phi_{\downarrow}(\mathbf{k})\rangle=\frac{1}{\sqrt{M^-_{\mathbf{k}}}}\begin{pmatrix}
        E_g-U-2\lambda_v\\
        -2tg^*
    \end{pmatrix}.
\end{equation}

We then find the Chern number of the negative energy state of the auxiliary Hamiltonian $H(\mathbf{k})=|\phi_{\downarrow}(\mathbf{k})\rangle\langle\phi_{\downarrow}(\mathbf{k})|-|\phi_{\uparrow}(\mathbf{k})\rangle\langle\phi_{\uparrow}(\mathbf{k})|$, which gives us $\nu_{\uparrow}=0$, in agreement with $(-1)^\nu=(-1)^{\nu_\uparrow}$. 

Note that our formalism can be applied even if a component of the spin is not conserved. In that case, care must be taken to ensure that the states $\ket{A,\sigma(\pmb\alpha)}$ are nonvanishing (i.e. normalizable) and pairwise linearly independent for all $\pmb{\alpha}$. This is a many-body analogy of the observation in Refs.~\cite{prodan2009robustness,lin2024spin} that the spin Chern number can only be computed for a noninteracting system if there is a gap in the spectrum of the projected spin operator.

\section{Conclusion}\label{sec:conclusion} 

\begin{table*}[ht]
\begin{tabular}{|c|c|}
\hline
 Non-interacting formula 
 
 & Many-body formula \\
 \hline\hline
 
 $\frac{1}{2\pi}[\oint_{\text{HBZ}}\mathbf{A}(\mathbf{k})-\int_{\text{HBZ}}F_{xy}(\mathbf{k})]\mod 2$ 
 
 & {\bf Eq.~\eqref{Z2_inv}} \\
 \hline
 
 $1/2$(Spin Chern number) $\mod 2$ 
 
 & { Chern number of the $|A,\sigma\rangle$ states}\\
 \hline

$\prod_{i=1}^4\frac{\sqrt{\det[w(\Gamma_i)]}}{\text{Pf}[w(\Gamma_i)]}$ ($\Gamma_i$'s are the TRIM points in the Brillouin zone) & The formula in Ref. \cite{leeManyBodyGeneralizationTopological2008}  \\
 \hline

  $\oint_{\text{HBZ}}\nabla \text{Pf}(m(\mathbf{k}))$(mod 2)
  
  & ?  \\
 \hline
\end{tabular}
\caption{Different formulations of the $\mathbb{Z}_2$ invariant for interacting and non-interacting systems.}\label{z2tab}
\end{table*}

In this work, we have derived a new expression Eq.~\eqref{Z2_inv} for the $\mathbb{Z}_2$ invariant of a many-body system in terms of the Berry connection of QSP states on the flux Brillouin zone. 
Our formula generalizes the results of Ref.~\cite{leeManyBodyGeneralizationTopological2008}, and connects the many-body $\mathbb{Z}_2$ invariant with the NTW expression Eq.~\eqref{eq:24} for the Chern number, analogous to the connection drawn in Ref.~\cite{fuTimeReversalPolarization2006} in the noninteracting case. 
Furthermore, our formulation of the invariant is well-suited for explicit calculations in orbital Hatsugai-Kohmoto models. 
We also show how the QSP states can be used to define the $\mathbb{Z}_2$ invariant through the spin Chern number. 
We have calculated the Chern number for the Haldane-HK model, as well as the $\mathbb{Z}_2$ invariant for the Kane-Mele model with orbital HK interaction. 
As far as we are aware, this represents the first direct analytic calculation of the $\mathbb{Z}_2$ invariant from an interacting many-body ground state, which demonstrates the utility of our approach. 

In Table~\ref{z2tab} we summarize the different formulations of the $\mathbb{Z}_2$ invariant now known. 
The left column gives the four known equivalent formulations of the invariant in terms of, respectively, an integral of the Berry connection over the HBZ~\cite{fuTimeReversalPolarization2006}, the parity of (half) the spin Chern number~\cite{prodan2009robustness,sheng2006quantum}, the phase of the eigenvalues of the sewing matrix for time-reversal symmetry~\cite{fuTimeReversalPolarization2006}, and zeros of the Pfaffian of the time-reversal matrix~\cite{kaneTopologicalOrderQuantum2005}. 
In the right hand column, we give the known extension of these formulas to many-body systems. 
Ref.~\cite{leeManyBodyGeneralizationTopological2008} had previously derived a formula for the invariant in terms of the many-body sewing matrix for time-reversal, and in terms of the spin Chern number of the QSP states. 
The bold entry represent the main results of this work. 
We have also established that these three forms of the many-body invariant are equivalent.

However, when it comes to the original formulation of the $\mathbb{Z}_2$ invariant in terms of zeros of the Pfaffian of the time-reversal matrix, we run into a problem. 
One can attempt to derive a many-body version of this expression by looking at the Pfaffian of the time-reversal matrix 
\begin{equation}\label{Eq:manybodym}
    m_{\sigma,\sigma'}(\pmb{\alpha})=\langle A,\sigma(\pmb{\alpha})|\Theta|A,\sigma(\pmb{\alpha})\rangle
\end{equation}
defined for QSP states in the flux BZ. 
However, we find that $m(\pmb{\alpha})$ defined in this way is not invariant under shifts of $\pmb{\alpha}$ by a flux Brillouin zone lattice vector. 
As such, it is not clear that the (parity) of the winding of Eq.~\eqref{Eq:manybodym} about its zeros is gauge invariant. 
Thus, we cannot directly use Eq.~\eqref{Eq:manybodym} to write the $\mathbb{Z}_2$ invariant. 
We leave the question of a Pfaffian formula along these lines for future work.

Next, we have focused in this work on systems with non-degenerate ground states. 
For systems with topological ground state degeneracy, the domain of the flux Brillouin zone needs to be extended. 
Ref.~\cite{niuQuantizedHallConductance1985} established how, for fractional quantum Hall states, the NTW formula \eqref{eq:24} can be extended to an average over the extended flux Brillouin zone to obtain fractionalized invariants. 
It is worthwhile to ask how one can compute topological invariants for fractionalized phases with time reversal symmetry. 
We believe that our Eq.~\eqref{Z2_inv} represents a well-founded starting point for such an investigation. 

Additionally, the application of the theory of symmetry indicators to the computation of \eqref{Z2_inv} is a fruitful area for future work. 
In systems with crystal symmetries (in particular inversion symmetry), One can follow the logic of Refs.~\cite{fu2007topological,po2017symmetrybased,fang2012bulk,bradlyn2017topological} to attempt to derive crystal symmetry constraints on the Berry curvature of the QSP states as a function of flux. 
For this to work, it is important that the choice of QSP state is consistent with the crystal symmetry: in the Kane-Mele model, for instance, the $\ket{A,\sigma(\pmb{\alpha})}$ states map into the $\ket{B,\sigma(\pmb{\alpha})}$ states under inversion symmetry, which is undesirable. 
However, for other choices of QSP states or for more general models (such as the square-lattice BHZ model~\cite{bernevig2006quantum}), the symmetry-indicator approach may be applicable.

Although we focused here on the topological invariants for two-dimensional topological insulators, our approach should also be applicable to three dimensional systems. 
We expect that the three-dimensional $\mathbb{Z}_2$ invariant can be expressed as the difference of Eq.~\eqref{Z2_inv} evaluated on two time-reversal invariant planes of the three-dimensional flux Brillouin zone, in analogy with the noninteracting invariant~\cite{fu2007topologicalinv,qi2008topological,yu2011equivalent,moore2007topological}. 
We defer a complete exploration of the three dimensional case to future work.

Finally, when it comes to HK models specifically, our work fills a gap in the literature by showing a direct computation of the topological invariant at nonzero $U$. 
However, it is an open question how to relate these invariants to observables in long-range interacting systems such as HK models. 
For instance,  we know that for short-range Hamiltonians a nonzero value of $\nu$ in Eq.~\eqref{Z2_inv} requires the presence of gapless helical excitations at a boundary via the bulk-boundary correspondence. 
For HK models, however, the bulk-boundary correspondence may fail as edge modes on opposite sides of the system will be nontrivially coupled. 
Similarly, for Chern insulators with HK interactions, chiral edge modes may not be present due to a failure of the bulk boundary correspondence. 
Our formulation of the $\mathbb{Z}_2$ invariant and Chern number in terms of QSP states gives a starting point for a more systematic study of the bulk-boundary correspondence in these systems, which would be a fruitful area for further study.

\begin{acknowledgments}
The authors thank Dmitry Manning-Coe and Jinchao Zhao for useful discussions. 
This work was supported by the U.S. 
DOE, Office of Basic Energy Sciences, Energy Frontier Research Center for Quantum Sensing and Quantum Materials through Grant No. 
DE-SC0021238. 
B.~B. received addition support from the Alfred P. 
Sloan Foundation and the National Science Foundation under grant DMR-1945058.
\end{acknowledgments}

\onecolumngrid

\appendix
\section{Proof of the TKNN formula}
\label{app:A}
To derive the TKNN formula~\cite{thoulessQuantizedHallConductance1982} from the NTW formula for a non-interacting ground state, we start with the single-particle Bloch Hamiltonian,
\begin{equation}
    \label{eq:25}
    H=\sum_{\mathbf{k}\in\text{BZ}}\sum_{\mu\mu'}c^{\dagger}_{\mathbf{k}\mu}[h(\mathbf{k})]_{\mu\mu'}c_{\mathbf{k}\mu'}.
\end{equation}
We add a flux to this Hamiltonian by replacing,
\begin{equation}
    \label{eq:26}
    H(\pmb{\alpha})=\sum_{\mathbf{k}\in\text{BZ}}\sum_{\mu\mu'}c^{\dagger}_{\mathbf{k}\mu}[h(\mathbf{k}-\pmb{\alpha})]_{\mu\mu'}c_{\mathbf{k}\mu'}.
\end{equation}
To find the ground state of Eq.~\eqref{eq:26}, we start by introducing a set of annihilation operators
\begin{equation}
    \Tilde{c}_{\mathbf{k}n}=\sum_{\mu}U^*_{\mu n}(\mathbf{k})c_{\mathbf{k}\mu},
\end{equation}
where $U_{\mu n}(\mathbf{k})$ is the matrix of eigenvectors of $h(\mathbf{k})$ (the matrix whose columns are $|u_n(\mathbf{k})\rangle$'s which satisfy $h(\mathbf{k})|u_n(\mathbf{k})\rangle=\epsilon_n(\mathbf{k})|u_n(\mathbf{k}\rangle$). 
In terms of $\Tilde{c}_{\mathbf{k}n}$ the Hamiltonian \eqref{eq:25} becomes,
\begin{equation}
    H=\sum_{\mathbf{k}n}\epsilon_n(\mathbf{k})\Tilde{c}^{\dagger}_{\mathbf{k}n}\Tilde{c}_{\mathbf{k}n}.
\end{equation}
However, this does not work for \eqref{eq:26} since $U_{\mu n}(\mathbf{k})$ does not diagonalize $h(\mathbf{k}-\pmb{\alpha})$. 
Instead, we define,
\begin{equation}
\label{eq:29}
    \Tilde{c}_{\mathbf{k}n}(\pmb{\alpha})=\sum_{\mu}U^*_{\mu n}(\mathbf{k}-\pmb{\alpha})c_{\mathbf{k}\mu}.
\end{equation}
These annihilation operators satisfy the usual anticommutation relations, and they diagonalize Eq.~\eqref{eq:26}. 
The ground state wavefunction for $N$ filled bands is then given by the Slater determinant
\begin{equation}
\label{eq:A6}    |\psi_0(\pmb{\alpha})\rangle=\prod_{\mathbf{k}\in\text{BZ}}\prod_{n=1}^N\Tilde{c}^{\dagger}_{\mathbf{k}n}(\pmb{\alpha})|0\rangle.
\end{equation}
IWe see that this ground state \eqref{eq:A6} satisfies normal periodic boundary conditions (no need to consider magnetic BCs since $B=0$), because the operators $\Tilde{c}^{\dagger}_{\mathbf{k}n}(\pmb{\alpha})$'s are a linear combination of $c^{\dagger}_{\mathbf{k}\mu}$'s which create the single-particle states $|\psi_{\mathbf{k}\mu}\rangle$ that are periodic when $\mathbf{x}\rightarrow\mathbf{x}+\mathbf{L}$.

One can shift $\pmb{\alpha}$ by a flux Brillouin zone vector, $\mathbf{q}=\frac{2\pi n_x}{L_x}\hat{\mathbf{x}}+\frac{2\pi n_y}{L_y}\hat{\mathbf{y}}$, and this is equivalent to a {\it local} gauge transformation of the many-body Hamiltonian, $H(\pmb{\alpha}+\mathbf{q})= {e}^{-{i}\mathbf{q}\cdot\mathbf{R}}H(\pmb{\alpha}){e}^{{i}\mathbf{q}\cdot\mathbf{R}}$, where $\mathbf{R}=\sum_{\mathbf{r},\mu}\mathbf{r}c^{\dagger}_{\mathbf{r}\mu}c_{\mathbf{r}\mu}$. 
This is because,
\begin{flalign}
    &c_{\mathbf{k}\mu}=\sum_{\mathbf{r}}{e}^{{i}\mathbf{k}\cdot\mathbf{r}}c_{\mathbf{r}\mu}\\&\implies{e}^{-{i}\mathbf{q}\cdot\mathbf{R}}c_{\mathbf{k}\mu}{e}^{{i}\mathbf{q}\cdot\mathbf{R}}=\sum_{\mathbf{r}}{e}^{{i}\mathbf{k}\cdot\mathbf{r}}{e}^{-{i}\mathbf{q}\cdot\mathbf{R}}c_{\mathbf{r}\mu}{e}^{{i}\mathbf{q}\cdot\mathbf{R}}\\&=\sum_{\mathbf{r}}{e}^{{i}(\mathbf{k}+\mathbf{q})\cdot\mathbf{r}}c_{\mathbf{r}\mu }=c_{(\mathbf{k}+\mathbf{q})\mu}.
\end{flalign}
This equation only makes sense when $\mathbf{q}$ is a flux Brillouin zone vector, otherwise, the annihilation operator does not satisfy the periodic boundary conditions. 
Similarly,
\begin{equation}
    {e}^{-{i}\mathbf{q}\cdot\mathbf{R}}c^{\dagger}_{\mathbf{k}\mu}{e}^{{i}\mathbf{q}\cdot\mathbf{R}}=c^{\dagger}_{(\mathbf{k}+\mathbf{q})\mu}.
\end{equation}
Therefore, the many-body Hamiltonian changes according to,
\begin{equation}
    H(\pmb{\alpha}+\mathbf{q})= {e}^{-{i}\mathbf{q}\cdot\mathbf{R}}H(\pmb{\alpha}){e}^{{i}\mathbf{q}\cdot\mathbf{R}}.
\end{equation}
This is accompanied by the following change in the ground state,
\begin{equation}
    |\psi_0(\pmb{\alpha}+\mathbf{q})\rangle={e}^{-{i}\mathbf{q}\cdot\mathbf{R}}|\psi_0(\pmb{\alpha})\rangle.
\end{equation}
Since this is a local gauge transformation, we restrict $\pmb{\alpha}$ to a torus of size $\frac{2\pi}{L_x}\times\frac{2\pi}{L_y}$,
\begin{equation}
    |\psi_0(\pmb{\alpha}+\mathbf{q})\rangle\sim|\psi_0(\pmb{\alpha})\rangle
\end{equation}
for any flux Brillouin zone vector $\mathbf{q}$.

If the Hamiltonian Eq.~\eqref{eq:26} has an energy gap for all choices of $\alpha$, we can adiabatically thread a flux through the system (twist the boundary conditions) without inducing a transition to any excited states.
In the thermodynamic limit, this lets us relate $\ket{\psi_0(\pmb{\alpha})}$ to $\ket{\psi_0(\pmb{\alpha}+\mathbf{q})}$ using adiabatic perturbation theory. 
For large $L$, $|\mathbf{q}|$ is small, and so we can expand
\begin{equation}
    H(\pmb{\alpha}+\mathbf{q})\sim H(\pmb{\alpha})+\mathbf{q}\cdot\nabla_{\alpha}H(\pmb{\alpha}).
\end{equation}
Using perturbation theory, we find to lowest order in $\mathbf{q}$ that
\begin{equation}
    |\psi_0(\pmb{\alpha}+\mathbf{q})\rangle\sim{e}^{{i}\gamma}\left(|\psi_0(\pmb{\alpha})\rangle+\sum_{j\neq 0}\frac{\langle\psi_j(\pmb{\alpha})|\mathbf{q}\cdot\nabla_{\alpha}H(\pmb{\alpha}|\psi_0(\pmb{\alpha})\rangle}{E_j(\pmb{\alpha})-E_0(\pmb{\alpha})}\right),
\end{equation}
where $\gamma$ is a (Berry) phase determined from the diagonal matrix element of $e^{-i\mathbf{q}\cdot\mathbf{R}}$. 
Plugging the new ground state \eqref{eq:A6} into the NTW formula \eqref{eq:24}, we find for the Chern number that
\begin{flalign}
    \bar{\sigma}_{xy}&=-\frac{e^2}{h}\frac{1}{2\pi{i}}\int_{-\pi/L_x}^{\pi/L_x}\int_{-\pi/L_y}^{\pi/L_y}\mathrm{d}^2\pmb{\alpha}\left(\braket{\frac{\partial\psi_0}{\partial\alpha_x}}{\frac{\partial\psi_0}{\partial\alpha_y}}-\braket{\frac{\partial\psi_0}{\partial\alpha_y}}{\frac{\partial\psi_0}{\partial\alpha_x}}\right)\nonumber\\
    &=-\frac{e^2}{h}\frac{1}{2\pi{i}}\sum_{\mathbf{k}\in\text{BZ}}\int\mathrm{d}^2\pmb{\alpha}\sum_{n=1}^N\left(\langle0|\frac{\partial\Tilde{c}_{\mathbf{k}n}(\pmb{\alpha})}{\partial\alpha_x}\frac{\partial\Tilde{c}^{\dagger}_{\mathbf{k}n}(\pmb{\alpha})}{\partial\alpha_y}|0\rangle-\langle0|\frac{\partial\Tilde{c}_{\mathbf{k}n}(\pmb{\alpha})}{\partial\alpha_y}\frac{\partial\Tilde{c}^{\dagger}_{\mathbf{k}n}(\pmb{\alpha})}{\partial\alpha_x}|0\rangle\right)\nonumber\\
    &=-\frac{e^2}{h}\frac{1}{2\pi{i}}\sum_{\mathbf{k}\in\text{BZ}}\int\mathrm{d}^2\pmb{\alpha}\sum_{\mu n}\left(\frac{\partial U^*_{\mu n}(\mathbf{k}-\pmb{\alpha})}{\partial\alpha_x}\frac{\partial U_{\mu n}(\mathbf{k}-\pmb{\alpha})}{\partial\alpha_y}-\frac{\partial U^*_{\mu n}(\mathbf{k}-\pmb{\alpha})}{\partial\alpha_y}\frac{\partial U_{\mu n}(\mathbf{k}-\pmb{\alpha})}{\partial\alpha_x}\right)\nonumber\\
    &=-\frac{e^2}{h}\frac{1}{2\pi{i}}\sum_{\mathbf{k}\in\text{BZ}}\int\mathrm{d}^2\pmb{\alpha}\sum_{\mu n}\left(\frac{\partial U^*_{\mu n}(\mathbf{k}-\pmb{\alpha})}{\partial k_x}\frac{\partial U_{\mu n}(\mathbf{k}-\pmb{\alpha})}{\partial k_y}-\frac{\partial U^*_{\mu n}(\mathbf{k}-\pmb{\alpha})}{\partial k_y}\frac{\partial U_{\mu n}(\mathbf{k}-\pmb{\alpha})}{\partial k_x}\right)\nonumber\\
    &=-\frac{e^2}{h}\frac{1}{2\pi{i}}\int_{\text{CBZ}}\mathrm{d}^2\mathbf{k}\sum_{n=1}^N\left(\braket{\frac{\partial u_n(\mathbf{k})}{\partial k_x}}{\frac{\partial u_n(\mathbf{k})}{\partial k_y}}-\braket{\frac{\partial u_n(\mathbf{k})}{\partial k_y}}{\frac{\partial u_n(\mathbf{k})}{\partial k_x}}\right),\label{eq:tknn}
\end{flalign}
where in the last step, we combined the sum over $\mathbf{k}$ and the integral over $\pmb{\alpha}$ into one continuous sum over the entire Brillouin zone (CBZ=continuous Brillouin zone). 
Eq.~\eqref{eq:tknn} is nothing other than the TKNN formula for the Hall conductance of a noninteracting system. 

\section{Equivalence of formula \eqref{Z2_inv} to the $\mathbb{Z}_2$ invariant in \cite{leeManyBodyGeneralizationTopological2008}}
\label{app:B}
We will now connect our formula \eqref{Z2_inv} for the $\mathbb{Z}_2$ to the formulation used in Ref.~\cite{leeManyBodyGeneralizationTopological2008}. 
There, the authors defined,
\begin{equation}\label{eq:leeinvariant}
    (-1)^{\nu}=\prod_{i=1}^4\frac{\sqrt{\det[w(\Gamma_i)]}}{\text{Pf}[w(\Gamma_i)]},
\end{equation}
$\Gamma_i$ being the four TRIM points in the flux Brillouin zone, and $w[\pmb{\alpha}]$ is the $U(2)$ sewing matrix for the time reversal operator,
\begin{equation}
    \Theta|A,\sigma(\pmb{\alpha})\rangle=\sum_{\sigma'}|A,\sigma'(-\pmb{\alpha})\rangle w_{\sigma,\sigma'}[\pmb{\alpha}].
\end{equation}
The sewing matrix satisfies $w[\pmb{\alpha}]^T=-w[-\pmb{\alpha}]$ by virtue of the gauge choice \eqref{eq:gaugechoice}.  
The derivation of Eq.~\eqref{eq:leeinvariant} from Eq.~\eqref{Z2_inv} is similar to the analogous derivation in the noninteracting case outlined in the appendix of \cite{fuTimeReversalPolarization2006} and \cite{yu2011equivalent}. 
We start with the Wilson loop,
\begin{equation}
    W(\alpha_y)=P\exp\left[{i}\oint A_x(\alpha_x,\alpha_y)\mathrm{d}\alpha_x\right],
\end{equation}
where $P$ denotes path ordering, and the line integral is calculated from $\alpha_x=-\pi/L_x$ to $\pi/L_x$ and $A_{\mu}(\pmb{\alpha})$ is the non-Abelian $U(2)$ Berry curvature computed from the QSP states. 
We first split the Berry connection into its trace ($U(1)$) and trace-free parts ($'sU(2)$),
\begin{equation}
    A_{\mu}^{mn}=\frac{1}{2}(a_{\mu}\delta^{mn}+A_{\mu}^i\sigma^{mn}_i).
\end{equation}
We do the same with the sewing matrix, $w[\pmb{\alpha}]$
\begin{equation}
    w[\pmb{\alpha}]={e}^{{i}\zeta(\pmb{\alpha})}\tilde{w}[\pmb{\alpha}].
\end{equation}
Then for the $'sU(2)$ part of the Berry connection, we have,
\begin{equation}
    A_{\mu}^i(-\pmb{\alpha})\sigma_i=A_{\mu}^i(\pmb{\alpha})\tilde{w}[\pmb{\alpha}]\sigma_i^T\tilde{w}[\pmb{\alpha}]^{\dagger}-2\tilde{w}[\pmb{\alpha}]\partial_{\mu}\tilde{w}[\pmb{\alpha}]^{\dagger}.
\end{equation}
We now decompose the Wilson loop at $\alpha_y^*=0,\pi/L_y$,
\begin{equation}
    W(\alpha_y^*)={e}^{{i}/2\oint a_x(\alpha_x,\alpha_y^*)\mathrm{d}\alpha_x}P\exp\left[{i}\oint \tilde{A}_x(\alpha_x,\alpha_y^*)\mathrm{d}\alpha_x\right],
\end{equation}
where $\tilde{A}_{\mu}=\frac{1}{2}A_{\mu}^i\sigma^{mn}_i$. 
Inside the path ordered exponential, we can write the integrand as,
\begin{equation}
    \int^{\pi/L_x}_{-\pi/L_x}\tilde{A}_x(\alpha_x,\alpha_y^*)=\int^{\pi/L_x}_0\tilde{A}_x(\alpha_x,\alpha_y^*)\mathrm{d}\alpha_x+\int^0_{-\pi/L_x}\tilde{A}_x(\alpha_x,\alpha_y^*)\mathrm{d}\alpha_x.
\end{equation}
We then use the path ordering to factorize the Wilson loop into two parts and rewrite,
\begin{flalign}
    \int^0_{-\pi/L_x}\tilde{A}_x(\alpha_x,\alpha_y^*)\mathrm{d}\alpha_x&=\int^{\pi/L_x}_0\tilde{A}_x(-\alpha_x,\alpha_y^*)\mathrm{d}\alpha_x\nonumber\\&=\int_0^{\pi/L_x}\left(\tilde{w}[\alpha_x,\alpha_y^*]\tilde{A}^T_x(\alpha_x,\alpha_y^*)\tilde{w}[\alpha_x,\alpha_y^*]^{\dagger}-\tilde{w}[\alpha_x,\alpha_y^*]\partial_{x}\tilde{w}[\alpha_x,\alpha_y^*]^{\dagger}\right)\mathrm{d}\alpha_x,
\end{flalign}
where we have used that $\tilde{w}[\alpha_x,-\alpha_y^*]=\tilde{w}[\alpha_x,\alpha_y^*]$ since $\tilde{w}$ is invariant under shifts of $\pmb{\alpha}$ by a Brillouin zone vector. 
After plugging this expression into the path ordered exponential, and noting that this is just a gauge transformation of $A_x^T$, we get,
\begin{equation}
    W(\alpha_y^*)={e}^{{i}/2\oint a_x(\alpha)\mathrm{d}\alpha_x}P\exp\left[{i}\int \tilde{A}_x(\alpha_x)\mathrm{d}\alpha_x\right]\tilde{w}[\pi/L_x]P\exp\left[{i}\int \tilde{A}^T_x(\alpha_x)\mathrm{d}\alpha_x\right]\tilde{w}[0]^{\dagger}.
\end{equation}
Since at the TRIM points, $\tilde{w}$ is proportional to ${i}\sigma_y$ (because of our gauge choice), we can use the anticommutation relation $\sigma_y\sigma^T_i=-\sigma_i\sigma_y$,  to exchange the third and fourth terms. 
We then see that the Wilson lines cancel and we are left with,
\begin{equation}
    W(\alpha_y^*)={e}^{{i}/2\oint a_x(\alpha_x,\alpha_y^*)\mathrm{d}\alpha_x}\text{Pf}(\tilde{w}[\pi/L_x,\alpha_y^*])\text{Pf}(\tilde{w}[0,\alpha_y^*]^{\dagger})\times\mathrm{Id},
\end{equation}
where we have replaced the sewing matrices with their Pfaffians, because at the TRIM points they are antisymmetric 2$\times$2 matrices. 
Therefore,
\begin{flalign}
\label{eq:95}
    & W(\pi/L_y)W(0)^{-1}\\
    &=\exp\left[{i}/2\oint_{\partial\text{HFBZ}}\mathrm{d}\pmb{\alpha}\cdot\mathbf{a}(\pmb{\alpha})\right]\prod_{i=1}^{4}\text{Pf}(\tilde{w}[\gamma_i]).
\end{flalign}
However, since the Wilson loops on the first line are proportional to $\sigma_y$, we can write 
\begin{equation}
    W(\pi/L_y)W(0)^{-1}=\exp\left[\int_0^{\pi/L_y}\nabla\log\sqrt{\det W(\alpha_y)}\right]=\exp\left[\frac{1}{2}\int_0^{\pi/L_y}\nabla\log\det W(\alpha_y)\right].
\end{equation}
Using the Ambrose Singer theorem, we can rewrite,
\begin{equation}
    {i}\nabla\log\det W(\alpha_y)=\int_{-\pi/L_x}^{\pi/L_x}F_{12}(\alpha_x,\alpha_y)\mathrm{d}\alpha_x,
\end{equation}
where $F_{12}$ is the curvature of the $U(1)$ part of the Berry connection. 
Plugging this into \eqref{eq:95}, we see that our $\mathbb{Z}_2$ invariant is equivalent to the one in \cite{leeManyBodyGeneralizationTopological2008}.

\section{Chern number in the Haldane-HK model}
\label{app:C}
In this Appendix, we will use the formulation Eq.~\eqref{eq:hkntw} of the NTW formula for HK models to compute the Chern number for the square lattice Chern insulator with HK interaction. 
We take as our starting point the Hamiltonian
\begin{align}
H&=H_0+H_U \\
H_0&=\sum_{\mathbf{k},\tau\tau'\sigma} c^\dag_{\mathbf{k}\tau\sigma}\mathbf{d}(\mathbf{k})\cdot\vec{\tau}_{\tau\tau'}c_{\mathbf{k}\tau'\sigma}-\mu_0\sum_{\mathbf{k}\tau\sigma} n_{\mathbf{k}\tau\sigma} \\
H_U&= U\sum_{\mathbf{k}\tau}n_{\mathbf{k}\tau\uparrow}n_{\mathbf{k}\tau\downarrow}.
\end{align}
Here $\vec{\tau}$ is a vector of Pauli matrices acting in a $2\times 2$ subspace of orbital degrees of freedom, $c_{\mathbf{k}\tau\sigma}$ creates an electron with spin $\sigma=\uparrow,\downarrow$ in orbital $\tau=1,2$ with momentum $\mathbf{k}$, $n_{\mathbf{k}\tau\sigma}=c^\dag_{\mathbf{k}\tau\sigma}c_{\mathbf{k}\tau\sigma}$, and $\mu_0$ is the chemical potential. 
The vector $\mathbf{d}(\mathbf{k})$ defines the single particle Hamiltonian; while much of what we will say will hold for general $\mathbf{d}$, we can focus on the particular case of a Chern insulator with
\begin{align}
d_x(\mathbf{k}) &= \sin k_x \\
d_y(\mathbf{k}) &= \sin k_y \\
d_z(\mathbf{k}) &= 2+m-\cos k_x-\cos k_y,
\end{align}
where $k_i\in[-\pi,\pi]$ is given in reduced coordinates. 
For this choice of $\mathbf{d}(\mathbf{k})$, the single particle Hamiltonian describes a QAH insulator with gapped Chern bands when $-4<m<0, m\neq -2$ and trivial bands when $m>0$ or $m<-4$. 
It will be important for our later analysis to recall that the Chern number $C$ of the lower band is given by the homotopy class of the unit vector $\hat{d}=\mathbf{d}/|\mathbf{d}|$ viewed as map from the Brillouin zone to the Bloch sphere,
\begin{equation}\label{eq:winding}
\nu=\frac{1}{8\pi}\int d^2k \epsilon^{\mu\nu}\hat{d}\cdot\left(\partial_\mu \hat{d}\times \partial_\nu\hat{d}\right).
\end{equation}
In particular, when $m>0$ or $m<-4$ the vector $\hat{d}(\mathbf{k})$ defines a map that is homotopic to the constant map, whereas when $-4<m<0$ the vector $\hat{d}(\mathbf{k})$ covers every point on the Bloch sphere.

Our goal will be to diagonalize this Hamiltonian. 
At each $\mathbf{k}$ point we can consider separately the sector of $N$-particle states. 
In the zero particle sector we have at every $\mathbf{k}$
\begin{equation}
E_{N=0}(\mathbf{k}) = 0.
\end{equation} 
In the one-particle sector, the interaction term has vanishing matrix elements and we have
\begin{equation}
E_{N=1}^{\pm}(\mathbf{k}) = \pm |\mathbf{d}(\mathbf{k})|-\mu_0.
\end{equation} 
Since the single-particle Hamiltonian is (up to a shift in chemical potential) invariant under the particle-hole transformation
\begin{equation}
Pc_{\tau\sigma}P^{-1} = \sum_{\tau'} \tau_y^{\tau\tau'}c_{\tau'\sigma},
\end{equation}
we can immediately deduce the energies in the three particle sector. 
They are
\begin{equation}
E_{N=3}^{\pm}(\mathbf{k}) =  \pm |\mathbf{d}(\mathbf{k})|-3\mu_0+U
\end{equation}
We note that when $\mu_0=U/2$ we have that $E_{N=3}^\pm=E_{N=1}^{\pm}$, consistent with particle-hole symmetry. 

We now focus on the $N=2$ sector. 
We choose the following ordered basis for the six $N=2$ states:
\begin{align}
\ket{11} &= c^\dag_{1\uparrow}c^\dag_{2\uparrow}\ket{0} \\
\ket{10} &= \frac{1}{\sqrt{2}}(c^\dag_{1\uparrow}c^\dag_{2\downarrow}+c^\dag_{1\downarrow}c^\dag_{2\uparrow})\ket{0} \\
\ket{1\bar{1}} &= c^\dag_{1\downarrow}c^\dag_{2\downarrow}\ket{0} \\
\ket{a} &= \frac{i}{\sqrt{2}}(c^\dag_{1\uparrow}c^\dag_{1\downarrow}+c^\dag_{2\uparrow}c^\dag_{2\downarrow})\ket{0} \\
\ket{b} &= \frac{1}{\sqrt{2}}(c^\dag_{2\uparrow}c^\dag_{2\downarrow}-c^\dag_{1\uparrow}c^\dag_{1\downarrow})\ket{0} \\
\ket{c} &= \frac{1}{\sqrt{2}}(c^\dag_{1\uparrow}c^\dag_{2\downarrow}-c^\dag_{1\downarrow}c^\dag_{2\uparrow})\ket{0} ,
\end{align}
The three states $\{\ket{11},\ket{10},  \ket{1\bar{1}}\}$ have total spin $'s=1$, while the three states $\{\ket{a},\ket{b},\ket{c}\}$ span the space of spin singlet states. 
The matrix of the Hamiltonian in this basis, which we denote as $H_2$, can be written in block form,
\begin{equation}\label{eq:h6}
H_2=\begin{pmatrix}
0_3 & 0_3 \\
0_3 & 2\mathbf{\tilde{d}}(\mathbf{k})\cdot\mathbf{L} + \frac{U}{\sqrt{3}}\lambda_8 +\frac{2U}{3}
\end{pmatrix} -2\mu_0.
\end{equation}
Here $0_3$ denotes a $3\times 3$ matrix of zeros, $\mathbf{\tilde{d}}=(-d_x,d_y,d_z)$ is given in terms of the vector $\mathbf{d}$ from the single particle Hamiltonian, $\mathbf{L}$ is a vector of spin-$1$ matrices in the vector representation, given explicitly by
\begin{align}
L_x &= \begin{pmatrix}
0 & 0 & -i \\
0 & 0 & 0 \\
i & 0 & 0
\end{pmatrix} \\
L_y &= \begin{pmatrix}
0 & 0 & 0 \\
0 & 0 & i \\
0 & -i & 0
\end{pmatrix} \\
L_z &= \begin{pmatrix}
0 & i & 0 \\
-i & 0 & 0 \\
0 & 0 & 0
\end{pmatrix},
\end{align}
and $\lambda_8$ is the Gell-Mann matrix
\begin{equation}
\lambda_8 = \frac{1}{\sqrt{3}}\begin{pmatrix}
1 & 0 & 0 \\
0 & 1 & 0 \\
0 & 0 & -2
\end{pmatrix}.
\end{equation}
We see that the three $'s=1$ states are trivial eigenvectors with energy $E_{N=2}^{S=1}=-2\mu_0$. 
The three nontrivial eigenstates $\ket{\psi_i}, i=1,2,3$ with energies $E_{N=2}^i$ span the spin singlet subspace. 
Writing
\begin{equation}
\ket{\psi_i} = u_i^a\ket{a}+ u_i^b\ket{b} u_i^c\ket{c}
\end{equation}
we have that $E_{N=2}^{(i)}$ and $\vec{u}_i$ are given as the eigenvalues and eigenvectors of the auxiliary $3\times 3$ matrix equation
\begin{equation}\label{eq:spin1}
\left(2\mathbf{\tilde{d}}(\mathbf{k})\cdot\mathbf{L} + \frac{U}{\sqrt{3}}\lambda_8 +\frac{2U}{3} - 2\mu_0\right)\vec{u}^i = E_{N=2}^{(i)}\vec{u}^i.
\end{equation}
The topology of eigenstates of spin-1 Hamiltonians such as this were considered in detail in Refs.~\cite{bradlyn2016dirac,flicker2018chiral}, where they emerge as the low-energy, single-particle $\mathbf{k}\cdot\mathbf{p}$ Hamiltonian near symmetry-protected degeneracies. 
Following Eq.~\eqref{eq:hkntw}, we can define the Chern number of the state $\ket{\psi_i}$ as
\begin{equation}
C_i = \frac{1}{2\pi i} \int d^2\mathbf{k}\left(\bra{\partial_x\psi_i}\ket{\partial_y\psi_i}-\bra{\partial_y\psi_i}\ket{\partial_x\psi_i}\right) 
\end{equation}
When $U=0$, we can use the analysis of Ref.~\cite{bradlyn2016dirac} to immediately deduce that the Chern number of the lowest-energy eigenstate of Eq.~\eqref{eq:spin1}, is given by \emph{twice} the winding of $\mathbf{d}$. 
Since the Chern number of the wave functions $\ket{\psi_i}$ cannot change unless a gap closes in the spectrum of Eq.~\eqref{eq:h6}. 
We thus expect the ground state Chern number to be given by $2\nu$ for small $U\neq 0$. 
For large $U\rightarrow\infty$, the ground state wavefunction is given by $\ket{c}$ plus small corrections, which has Chern number $0$. 
Thus, as a function of $U$ there is a topological phase transition.

We can go further and find an exact expression for the energies $E_{N=2}^{(i)}$. 
Letting $x=E+2\mu_0-2U/3$, we can write the characteristic polynomial corresponding to Eq.~\eqref{eq:spin1} as
\begin{equation}
P(x) = \det(2\mathbf{\tilde{d}}(\mathbf{k})\cdot\mathbf{L} + \frac{U}{\sqrt{3}}\lambda_8 -x) = -(x^3+px+q)
\end{equation}
where
\begin{align}
p &= -(4|\mathbf{d}|^2 + U^2/3) \\
q &= \frac{2U^3}{27} + \frac{4U}{3}(|d_x|^2+|d_y|^2-2|d_z|^2)
\end{align}
The roots of this depressed cubic can be given exactly in closed form (see, e.g., Appendix C of Ref.~\cite{flicker2018chiral}), yielding
\begin{equation}\label{eq:exactenergies}
E_{N=2}^{(i)} = 2U/3-2\mu_0 + 2\sqrt{\frac{-p}{3}}\cos\left[\frac{1}{3}\arccos\left(\frac{3q}{2p}\sqrt{\frac{-3}{p}}\right)-\frac{2\pi i}{3}\right].
\end{equation}
where we choose the principal branch $0\leq \arccos(x) \leq \pi$ of the inverse cosine. 
With this choice, the lowest energy is always given by $E_{N=2}^{(2)}$. 
Note also that we can use Eq.~\eqref{eq:exactenergies} to deduce there is a topological phase transition in the topological phase when $U=2|m|,\mu_0=U/2$. 
For these parameter values, the energy $E_{N=2}^{(2)}$ becomes equal to the energy $E^{S=1}_{N=2}$ of the spin triplet states at $k_x=k_y=0$, indicating a many-body gap closing and hence a topological phase transition to a metallic phase.
\section{Polarization in the Su-Schrieffer-Heeger-HK model}
\label{app:D}
In this Appendix we will be using the many-body formula for the polarization outlined in Ref. \cite{souza2000polarization} to compute the polarization of the Su-Schrieffer-Heeger (SSH) model \cite{sshmodel} with an orbital HK interaction. 
The polarization $P$ is given, mod the electric charge, by
\begin{flalign}\label{eq:polformula}
    P=\frac{ie}{2\pi}\int\mathrm{d}\alpha\langle\Phi(\alpha)|\partial_{\alpha}|\Phi(\alpha)\rangle\quad
    \text{(mod e)},
\end{flalign}
where $|\Phi(\alpha)\rangle$ is the ground state of the many-body Hamiltonian $H(\alpha)$ twisted by an angle $\alpha$ that satisfies periodic boundary conditions.

We will apply Eq.~\eqref{eq:polformula} to compute the polarization of the SSH model with an orbital HK interaction. The Hamiltonian for this model in the fully polarized (nontrivial) state is,
\begin{align}
    H=-t\sum_{k\sigma}(e^{ik}c^{\dagger}_{kA\sigma}c_{kB\sigma}+\text{h.c.})
+U\sum_{k\mu}n_{k\mu\uparrow}n_{k\mu\downarrow}.
\end{align}
Here, $A$ and $B$ label the sublattice sites in the one dimensional chain.
We will focus on the two-particle ground state of this Hamiltonian. 
The six basis states in this sector are given in Table \ref{basistable}.
In the two-particle sector, the Hamiltonian only has nonzero matrix elements between states $2,3,4,$ and $5$. The Hamiltonian matrix in this subspace takes the form
\begin{equation}
    H(k)=\begin{pmatrix}
U & te^{-i k} & te^{-i k} & 0 \\
te^{i k} & 0 & 0 & te^{-i k} \\
te^{i k} & 0 & 0 & te^{-i k} \\
0 & te^{i k} & te^{i k} & U 
\end{pmatrix}
\end{equation}
and its corresponding ground state eigenvector,
\begin{align}\label{eq:sshhkgndstate}
    U^{\mathrm{g.s.}}(k)=\frac{1}{\sqrt{N}}\begin{pmatrix}4te^{-ik}\\
        -U-\sqrt{U^2+16t^2} \\ -U-\sqrt{U^2+16t^2}\\4t e^{ik}
    \end{pmatrix}
\end{align}
where $N$ is a constant that is independent of k.

In terms of $U^{\mathrm{g.s.}}(k)$ the ground state can be written as
\begin{equation}
    |\Phi(\alpha)\rangle=\prod_{k\in\text{BZ}}\left(\sum_{\mu\mu'}U^{\mathrm{g.s.}}_{\mu\mu'}(k-\alpha)c^{\dagger}_{k\mu\uparrow}c^{\dagger}_{k\mu'\downarrow}\right)|0\rangle.
\end{equation}
Plugging this ground state into Eq.~\eqref{eq:polformula}, we get
\begin{equation}
    P=\frac{ie}{2\pi}\int \mathrm{d}k[U^{\mathrm{g.s.}}(k)]^*\partial_kU^{\mathrm{g.s.}}(k)\quad\text{(mod $e$)}.
\end{equation}

To evaluate this, we need to compute the Berry connection $i[U^{\mathrm{g.s.}}(k)]^*\partial_kU^{\mathrm{g.s.}}(k)$ associated to the ground state. Using Eq.~\eqref{eq:sshhkgndstate} we find
\begin{align*}
    i[U^{\mathrm{g.s.}}(k)]^*\partial_kU^{\mathrm{g.s.}}(k)=i\frac{1}{N}\begin{pmatrix}4t e^{-ik},
        -U-\sqrt{U^2+16t^2}, -U-\sqrt{U^2+16t^2},4te^{ik}
    \end{pmatrix}\cdot\begin{pmatrix}
        i4t e^{ik} \\ 0\\0\\-i4te^{-ik}
    \end{pmatrix}=0
\end{align*}
We an see that the polarization of the ground state is $0 \mod e$. 
This is the expected result since we know that in the limit $U\rightarrow 0$ each spin contributes $e/2 \mod e$ to the polarization, yielding a total polarization of $e \equiv 0 \mod e$.

In order to probe the topological structure of the ground state further, we can take inspiration from section \ref{many-body z2} and define many-body spin polarized states,
\begin{equation}
    |\uparrow,\mu(\alpha)\rangle\propto\left(\prod_{k\in\mathrm{BZ}}c_{k\mu\downarrow}\right)|\Phi(\alpha)\rangle,
\end{equation}
where $\mu$ is a sublattice index. 
%Unlike the quasi-single-particle states considered in the main text, these states are labelled by their spin.
We orthonormalize these states by considering the  linear combinations, 
\begin{align}
|n(\alpha)\rangle&=\sum_{\mu=A,B}U_{n\mu}(\alpha)|\uparrow,\mu(\alpha)\rangle,\nonumber \\
 \langle n(\alpha)|m(\alpha)\rangle&=\delta_{nm},\label{eq:sshortho}
\end{align}
with the additional gauge constraint from inversion symmetry [c.f. Eqs.~\eqref{TRstate1} and \eqref{TRstate2}],
\begin{equation}
 \label{constraint}   \mathbb{I}|n(\alpha)\rangle\propto|n(-\alpha)\rangle,
\end{equation}
where $\mathbb{I}$ is the inversion operator that acts on the annihilation operators as,
\begin{equation}
    \mathbb{I}^{-1}c_{\mathbf{k}A\sigma}\mathbb{I}=c_{-\mathbf{k}B\sigma}.
\end{equation}
Finally, we define,
\begin{equation}
    P_{\uparrow}=\frac{ie}{2\pi}\int\mathrm{d}\alpha\langle n=1,\uparrow(\alpha)|\partial_{\alpha}|n=1,\uparrow(\alpha)\rangle.
\end{equation}
The gauge constraint, \eqref{constraint} quantizes the value of $P_{\uparrow}$ to be 0 or $e/2$ (mod $e$).

Applying this to our ground state Eq.~\eqref{eq:sshhkgndstate} of the SSH model with HK interaction, we find that the QSP states are
\begin{equation}
    |\uparrow, A(\alpha)\rangle=\prod_{k\in\text{BZ}}\left[-(U+\sqrt{U^2+16t^2})c^{\dagger}_{kB\uparrow}-4te^{i(k-\alpha)}c^{\dagger}_{kA\uparrow}\right]|0\rangle
\end{equation}
\begin{equation}
    |\uparrow,B(\alpha)\rangle=\prod_{k\in\text{BZ}}\left[-(U+\sqrt{U^2+16t^2})c^{\dagger}_{kA\uparrow}-4te^{-i(k-\alpha)}c^{\dagger}_{kB\uparrow}\right]|0\rangle
\end{equation}
To satisfy the orthonormalization and gauge constraints in Eqs.~\eqref{eq:sshortho} and \eqref{constraint}, we define the linear combinations
\begin{equation}
    |1(\alpha)\rangle=\frac{1}{\sqrt{N_1}}|\uparrow,A(\alpha)\rangle+\frac{1}{\sqrt{N_1}}\Bigg(\prod_{k\in\text{BZ}}e^{i(k-\alpha)}\Bigg)|\uparrow,B(\alpha)\rangle
\end{equation}
\begin{equation}
    |2(\alpha)\rangle=\frac{1}{\sqrt{N_2}}|\uparrow,A(\alpha)\rangle-\frac{1}{\sqrt{N_2}}\Bigg(\prod_{k\in\text{BZ}}e^{i(k-\alpha)}\Bigg)|\uparrow,B(\alpha)\rangle,
\end{equation}
where $N_i=\prod_{k\in\text{BZ}}2\left(U+\sqrt{U^2+16t^2}-(-1)^i4t\right)^2=[2(U+\sqrt{U^2+16t^2}-(-1)^i4t)^2]^V$ and $V$ is the number of points in Brillouin zone. 
Plugging this into Eq.~\eqref{eq:polformula} for the polarization which only involves the $n=1$ states, and using that
\begin{align}
    \langle\uparrow,A(\alpha)|\partial_{\alpha}|\uparrow,A(\alpha)\rangle= i[2(4t+U\sqrt{U^2+16t^2})^2]^{V-1}\sum_{k\in\text{BZ}}16t^2
\end{align}
\begin{align}
    \langle\uparrow,A(\alpha)|\partial_{\alpha}\left[\Bigg(\prod_{k\in\text{BZ}}e^{i(k-\alpha)}\Bigg)|\uparrow,B(\alpha)\rangle\right]=i[2(4t+U\sqrt{U^2+16t^2})^2]^{V-1}\sum_{k\in\text{BZ}}4t(U+\sqrt{U^2+16t^2})
\end{align}
\begin{align}
    \langle\uparrow,B(\alpha)|\Bigg(\prod_{k\in\text{BZ}}e^{-i(k-\alpha)}\Bigg)\partial_{\alpha}|\uparrow,A(\alpha)\rangle=i[2(4t+U\sqrt{U^2+16t^2})^2]^{V-1} \sum_{k\in\text{BZ}}4t(U+\sqrt{U^2+16t^2})
\end{align}
\begin{align}
    \langle\uparrow,B(\alpha)|\Bigg(\prod_{k\in\text{BZ}}e^{-i(k-\alpha)}\Bigg)\partial_{\alpha}\left[\Bigg(\prod_{k\in\text{BZ}}e^{i(k-\alpha)}\Bigg)|\uparrow,B(\alpha)\rangle\right]=i[2(4t+U\sqrt{U^2+16t^2})^2]^{V-1} \sum_{k\in\text{BZ}}(U+\sqrt{U^2+16t^2})^2
\end{align}
we get,
\begin{equation}
    P_{\uparrow}=\frac{e}{2\pi}\int_{k\in\text{BZ}}1/2=\frac{e}{2},
\end{equation}
where we have combined the discrete sum over the $k$ and the integral over the flux Brillouin zone into an integral over the continuous Brillouin zone. 
Although we only looked at the spin up contribution to the total polarization, repeating the above steps for the down spin QSPs will give us the same answer mod $e$ (as is guaranteed from time-reversal symmetry).

\section{The many-body $\mathbb{Z}_2$ invariant for the Kane-Mele-HK model}
\label{app:E}
The ground states for Kane-Mele-HK model that we analyzed in Sec.~\ref{sec:kmhk_main} take the form
\begin{equation}
    |\psi_0(\pmb{\alpha})\rangle=\prod_{\mathbf{k}\in\text{BZ}}\left(\sum_{\mu\mu'}U^{\text{g.s.}}_{\mu\mu'}(\mathbf{k}-\pmb{\alpha})c^{\dagger}_{\mathbf{k}\mu\uparrow}c^{\dagger}_{\mathbf{k}\mu'\downarrow}\right)|0\rangle.
\end{equation}
The corresponding QSP states have the form
\begin{equation}\label{eq:appendix_berry}
    |A,\sigma(\pmb{\alpha})\rangle=\prod_{\mathbf{k}\in\text{BZ}}\left(U^{\sigma}_{\mu}(\mathbf{k}-\pmb{\alpha})c^{\dagger}_{\mathbf{k}\mu\sigma}\right)|0\rangle,
\end{equation}
where $U^{\sigma}_{\mu}(\mathbf{k})$'s are normalized, and there is no sum over $\sigma$. 
The corresponding gauge potential is,
\begin{equation}
    \mathbf{A}(\pmb{\alpha})=\sum_{\mathbf{k}\in\text{BZ}}\sum_{\mu\sigma}U^{\sigma}_{\mu}(\mathbf{k}-\pmb{\alpha})^*({i}\nabla_{\pmb{\alpha}}U_{\mu}^{\sigma}(\mathbf{k}-\pmb{\alpha})).
\end{equation}
For ease of calculation, we can introduce an auxiliary state
\begin{equation}\label{eq:auxstate}
    |\phi(\pmb{\alpha})\rangle=\prod_{\mathbf{k}\in\text{BZ}}\left(\sum_{\mu\mu'}V_{\mu\mu'}(\mathbf{k}-\pmb{\alpha})c^{\dagger}_{\mathbf{k}\mu\uparrow}c^{\dagger}_{\mathbf{k}\mu'\downarrow}\right)|0\rangle,
\end{equation}
where $V_{\mu\mu'}(\mathbf{k})=U^{\uparrow}_{\mu}(\mathbf{k})U^{\downarrow}_{\mu'}(\mathbf{k})$, which satisfies
\begin{equation}
    V_{\mu\mu'}(-\mathbf{k})=V^*_{\mu'\mu}(\mathbf{k})
\end{equation}
since $U^{\uparrow}_{\mu}(-\mathbf{k})=[U^{\downarrow}_{\mu}(\mathbf{k})]^*$. 
The gauge potential for the auxiliary state~\eqref{eq:auxstate} is given by
\begin{equation}\label{eq:auxberry}
    \mathbf{A}(\pmb{\alpha})=\sum_{\mathbf{k}\in\text{BZ}}\sum_{\mu\mu'}V_{\mu\mu'}(\mathbf{k}-\pmb{\alpha})^*\left({i}\nabla_{\pmb{\alpha}}V_{\mu\mu'}(\mathbf{k}-\pmb{\alpha})\right).
\end{equation}
Crucially, due to the normalization of the $U_\mu^\sigma(\mathbf{k})$'s, Eq.~\eqref{eq:auxberry} coincides with Eq.~\eqref{eq:appendix_berry}. 

We can split Eq.~\eqref{eq:auxberry} into two parts,
\begin{flalign}
    \mathbf{A}(\pmb{\alpha})&=\sum_{\mathbf{k}\in\text{BZ}}\sum_{\mu\mu'}V_{\mu\mu'}(\mathbf{k}-\pmb{\alpha})^*\left({i}\nabla_{\pmb{\alpha}}V_{\mu\mu'}(\mathbf{k}-\pmb{\alpha})\right)\\
    &=\sum_{\mathbf{k}\in\text{HBZ}}\sum_{\mu\mu'}V_{\mu\mu'}(\mathbf{k}-\pmb{\alpha})^*\left({i}\nabla_{\pmb{\alpha}}V_{\mu\mu'}(\mathbf{k}-\pmb{\alpha})\right)+\sum_{\mathbf{k}\in\overline{\text{HBZ}}}\sum_{\mu\mu'}V_{\mu\mu'}(\mathbf{k}-\pmb{\alpha})^*\left({i}\nabla_{\pmb{\alpha}}V_{\mu\mu'}(\mathbf{k}-\pmb{\alpha})\right)\\&=\sum_{\mathbf{k}\in\text{HBZ}}\sum_{\mu\mu'}V_{\mu\mu'}(\mathbf{k}-\pmb{\alpha})^*\left({i}\nabla_{\pmb{\alpha}}V_{\mu\mu'}(\mathbf{k}-\pmb{\alpha})\right)+\sum_{\mu\mu'}V_{\mu\mu'}(-\mathbf{k}-\pmb{\alpha})^*\left({i}\nabla_{\pmb{\alpha}}V_{\mu\mu'}(-\mathbf{k}-\pmb{\alpha})\right)\\&=\sum_{\mathbf{k}\in\text{HBZ}}\sum_{\mu\mu'}V_{\mu\mu'}(\mathbf{k}-\pmb{\alpha})^*\left({i}\nabla_{\pmb{\alpha}}V_{\mu\mu'}(\mathbf{k}-\pmb{\alpha})\right)+\sum_{\mu\mu'}V_{\mu'\mu}(\mathbf{k}+\pmb{\alpha})\left(-{i}\nabla_{-\pmb{\alpha}}\right)\left(V_{\mu'\mu}(\mathbf{k}+\pmb{\alpha})\right)^*\\&=\sum_{\mathbf{k}\in\text{HBZ}}\sum_{\mu\mu'}V_{\mu\mu'}(\mathbf{k}-\pmb{\alpha})^*\left({i}\nabla_{\pmb{\alpha}}V_{\mu\mu'}(\mathbf{k}-\pmb{\alpha})\right)+\sum_{\mu\mu'}V_{\mu\mu'}(\mathbf{k}+\pmb{\alpha})^*\left({i}\nabla_{-\pmb{\alpha}}V_{\mu\mu'}(\mathbf{k}+\pmb{\alpha})\right)\\&\label{A1+A2}=\mathbf{A}_1(\pmb{\alpha})+\mathbf{A}_1(-\pmb{\alpha}).
\end{flalign}
where $\mathbf{A}_1(\pmb{\alpha})=\sum_{\mathbf{k}\in\text{HBZ}}\mathbf{A}(\mathbf{k},\pmb{\alpha})$. 
The $\mathbf{A}(\mathbf{k},\pmb{\alpha})$'s are the gauge potentials of the normalized states, $V_{\mu\mu'}(\mathbf{k}-\pmb{\alpha})$. 
Note also that, by Stokes's theorem
\begin{equation}\label{eq:windingequation}
    \frac{1}{2\pi}\left[\oint_{\partial\text{HFBZ}}\mathrm{d}\pmb{\alpha}\cdot\mathbf{A}(\mathbf{k},\pmb{\alpha})-\int_{\text{HFBZ}}\mathrm{d}^2\pmb{\alpha}F_{xy}(\mathbf{k},\pmb{\alpha})\right]\in\mathbb{Z},
\end{equation}
since this is the difference between the line integral along a closed curve of $\mathbf{A}(\mathbf{k},\pmb{\alpha})$ evaluated in two different gauges; Eq.~\eqref{eq:windingequation} is the integer winding number of the gauge transformation that relates the two gauges. 
For the Berry curvature, we also have,
\begin{equation}
\label{F1+F2}
    F(\pmb{\alpha})=F_1(\pmb{\alpha})-F_1(-\pmb{\alpha}).
\end{equation}
Putting this all together, we have,
\begin{flalign}
    \nu&=\frac{1}{2\pi}\left[\oint_{\partial\text{HFBZ}}\mathrm{d}\pmb{\alpha}\cdot\mathbf{A}(\pmb{\alpha})-\int_{\text{HFBZ}}\mathrm{d}^2\pmb{\alpha}F_{xy}(\pmb{\alpha})\right]\label{eq:z2deriv1}\\
    &=\frac{1}{2\pi}\left[\oint_{\partial\text{HFBZ}}\mathrm{d}\pmb{\alpha}\cdot\mathbf{A}_1(\pmb{\alpha})-\int_{\text{HFBZ}}\mathrm{d}^2\pmb{\alpha}F_{1xy}(\pmb{\alpha})\right]+\frac{1}{2\pi}\left[\oint_{\partial{\text{HFBZ}}}\mathrm{d}\pmb{\alpha}\cdot\mathbf{A}_1(-\pmb{\alpha})+
    \int_{{\text{HFBZ}}}\mathrm{d}^2\pmb{\alpha}F_{1xy}(-\pmb{\alpha})\right]\label{eq:z2deriv2}\\
    &=\frac{1}{2\pi}\left[\oint_{\partial\text{HFBZ}}\mathrm{d}\pmb{\alpha}\cdot\mathbf{A}_1(\pmb{\alpha})-\int_{\text{HFBZ}}\mathrm{d}^2\pmb{\alpha}F_{1xy}(\pmb{\alpha})\right]-\frac{1}{2\pi}\left[\oint_{\partial{\overline{\text{HFBZ}}}}\mathrm{d}\pmb{\alpha}\cdot\mathbf{A}_1(\pmb{\alpha})-\int_{\overline{{\text{HFBZ}}}}\mathrm{d}^2\pmb{\alpha}F_{1xy}(\pmb{\alpha})\right]\label{eq:z2deriv3}\\
    &=\frac{1}{2\pi}\left[\oint_{\partial\text{FBZ}}\mathrm{d}\pmb{\alpha}\cdot\mathbf{A}_1(\pmb{\alpha})-\int_{\text{FBZ}}\mathrm{d}^2\pmb{\alpha}F_{1xy}(\pmb{\alpha})\right]-\frac{2}{2\pi}\left[\oint_{\partial\overline{\text{HFBZ}}}\mathrm{d}\pmb{\alpha}\cdot\mathbf{A}_1(\pmb{\alpha})-\int_{\overline{\text{HFBZ}}}\mathrm{d}^2\pmb{\alpha}F_{1xy}(\pmb{\alpha})\right]\label{eq:z2deriv4}.
\end{flalign}
We used Eq. \eqref{A1+A2} and Eq. \eqref{F1+F2} to get Eq.~\eqref{eq:z2deriv2}, changed integration variables from $\pmb{\alpha}$ to $-\pmb{\alpha}$ (the line integral picks up an extra negative sign when we change variables) to get Eq.~\eqref{eq:z2deriv3} and finally added 
 and subtracted $\frac{1}{2\pi}\left[\oint_{\partial\overline{\text{HFBZ}}}\mathrm{d}\pmb{\alpha}\cdot\mathbf{A}_1(\pmb{\alpha})-\int_{\overline{\text{HFBZ}}}\mathrm{d}^2\pmb{\alpha}F_{1xy}(\pmb{\alpha})\right]$ to arrive at Eq.~\eqref{eq:z2deriv4}. 
 Note that from Eq.~\eqref{eq:windingequation}, we see that the second term in Eq.~\eqref{eq:z2deriv4} is an even integer. 
 Therefore, modulo 2,
 \begin{flalign}
     \nu&=\frac{1}{2\pi}\left[\oint_{\partial\text{FBZ}}\mathrm{d}\pmb{\alpha}\cdot\mathbf{A}_1(\pmb{\alpha})-\int_{\text{FBZ}}\mathrm{d}^2\pmb{\alpha}F_{1xy}(\pmb{\alpha})\right]\nonumber\\
    &=\frac{1}{2\pi}\left[\oint_{\partial\text{HCBZ}}\mathrm{d}\mathbf{k}\cdot\mathbf{A}(\mathbf{k})-\int_{\text{HCBZ}}\mathrm{d}^2\mathbf{k}F_{xy}(\mathbf{k})\right]\mod 2,\label{eq:appZ2final}
\end{flalign}
where the integrals are over half of the continuous Brillouin zone (HCBZ),  
\begin{equation}
 \mathbf{A}(\mathbf{k})=\sum_{\mu\sigma}U_{\mu}^\sigma(\mathbf{k})^*\left({i}\nabla_{\mathbf{k}}U_{\mu}^\sigma(\mathbf{k})\right).
\end{equation}
is the Berry connection associated to the auxiliary QSP wavefunction $U_{\mu}^\sigma(\mathbf{k})$, and $F_{xy}(\mathbf{k})$ is its associated Berry curvature. 
Thus, for our models, we only need to compute the gauge potentials of the $U^{\sigma}_{\mu}(\mathbf{k})$'s and integrate them across half of the Brillouin zone. 
For non-interacting systems, the $U_{\mu}^{\sigma}$'s are the single-particle Bloch functions, so Eq.~\eqref{eq:appZ2final} reduces to the expression of Ref.~\cite{fuTimeReversalPolarization2006}. 
\twocolumngrid
\bibliography{MyLibrary.bib}

\end{document}